\documentclass[twocolumn]{aastex631}

\usepackage[utf8]{inputenc}
\usepackage{xcolor}
\newcommand{\uat}[2]{#1 (#2)}

\begin{document}

\title{Discovery of 51 pulsars in a reanalysis of the Green Bank North Celestial Cap Survey and a revised pulsar census}

\author[0000-0002-9409-3214]{R. Sengar}
\affiliation{Max Planck Institute for Gravitational Physics (Albert Einstein Institute), D-30167 Hannover, Germany \\}
\affiliation{Leibniz Universit\"{a}t Hannover, D-30167 Hannover, Germany\\}

\author[0000-0003-0292-4453]{L. Vleeschower}
\affiliation{Center for Gravitation, Cosmology, and Astrophysics, Department of Physics and Astronomy, University of Wisconsin-Milwaukee, P.O. Box 413, Milwaukee, WI 53201, USA\\}

\author[0000-0001-6295-2881]{D.~L. Kaplan}
\affiliation{Center for Gravitation, Cosmology, and Astrophysics, Department of Physics and Astronomy, University of Wisconsin-Milwaukee, P.O. Box 413, Milwaukee, WI 53201, USA\\}

\author[0000-0001-5134-3925]{G.~Y. Agazie}
\affiliation{Center for Gravitation, Cosmology, and Astrophysics, Department of Physics and Astronomy, University of Wisconsin-Milwaukee, P.O. Box 413, Milwaukee, WI 53201, USA\\}

\author[0000-0003-3294-3081]{M. Bailes}
\affiliation{Centre for Astrophysics and Supercomputing, Swinburne University of Technology, P.O. Box 218, Hawthorn, Victoria 3122, Australia\\}
\affiliation{Australian Research Council Centre of Excellence for Gravitational Wave Discovery (OzGrav)\\}

\author[0009-0007-0757-9800]{A. Cassity}
\affiliation{Department of Physics and Astronomy, University of British Columbia, 6224 Agricultural Road, Vancouver, BC V6T 1Z1, Canada\\}

\author[0000-0002-6039-692X]{H. T. Cromartie}
\affiliation{Department of Physics and Astronomy, Vanderbilt University, 2301 Vanderbilt Place, Nashville, TN 37235, USA\\}

\author[0000-0002-2185-1790]{M. E. Decesar}
\affiliation{Department of Physics and Astronomy, George Mason University, Fairfax, VA 22030, USA\\}

\author[0000-0001-8885-6388 ]{T. Dolch}
\affiliation{Department of Physics, Hillsdale College, 33 E. College Street, Hillsdale, MI 49242, USA\\}
\affiliation{Eureka Scientific, 2452 Delmer Street, Suite 100, Oakland, CA 94602-3017, USA\\}

\author[0000-0001-5645-5336]{W. Fiore}
\affiliation{Department of Physics and Astronomy, University of British Columbia, 6224 Agricultural Road, Vancouver, BC V6T 1Z1, Canada\\}

\author[0000-0001-8384-5049]{E. Fonseca}
\affiliation{Department of Physics and Astronomy, West Virginia University, P.O. Box 6315, Morgantown, WV 26506, USA\\}
\affiliation{Center for Gravitational Waves and Cosmology, West Virginia University, Chestnut Ridge Research Building, Morgantown, WV 26505, USA\\}

\author[0000-0001-5229-7430]{R. S. Lynch}
\affiliation{Green Bank Observatory, P.O. Box 2, Green Bank, WV 24944, USA\\}

\author[0000-0002-0430-6504]{E. Parent}
\affiliation{Université Grenoble Alpes, CNRS, IPAG, F-38000 Grenoble, France\\}

\author[0000-0001-5799-9714]{S. M. Ransom}
\affiliation{National Radio Astronomy Observatory, 520 Edgemont Road, Charlottesville, VA 22903, USA\\}

\author[0000-0001-9784-8670]{I. H. Stairs}
\affiliation{Department of Physics and Astronomy, University of British Columbia, 6224 Agricultural Road, Vancouver, BC V6T 1Z1, Canada\\}

\author[0000-0001-8503-6958]{J. van Leeuwen}
\affiliation{ASTRON, Netherlands Institute for Radio Astronomy, Oude Hoogeveensedijk 4, 7991 PD Dwingeloo, The Netherlands\\}
\affiliation{Leiden Observatory, Leiden University, Einsteinweg 55, 2333 CC Leiden, The Netherlands}
\affiliation{Leiden Institute for Advanced Computer Science (LIACS), Leiden University, Einsteinweg 55, 2333 CC Leiden, The Netherlands}

\begin{abstract}
We present results from a reanalysis of the Green Bank North Celestial Cap (GBNCC) survey, conducted at 350~MHz with the Green Bank Telescope between 2009 and 2020, covering the sky with declinations $\delta>-40^\circ$. We searched the survey data using an FFT-based periodicity search, with candidate selection extending to very low spectral signal-to-noise ratios and summing up to 32 harmonics. From the 60 promising candidates identified in this search, we confirmed 51 new pulsars using the Green bank and CHIME. We also used a direct folding search method (DFSM) to search for slow pulsars between 500~ms and 10~s. Of the 51 pulsars identified in the FFT search, 43 fall within this period range, all of which were found by the DFSM. The DFSM also found an additional pulsar that had been missed during the inspection of the FFT candidates. Most of the new discoveries are canonical pulsars, with spin periods ranging from 299~ms to 4.4~s and dispersion measures from 10 to 192~$\rm pc \, cm^{-3}$. We also found a binary millisecond pulsar with a spin period of 3.2~ms in a 4.5-d orbit. We redetected 855 pulsars in the data, 825 of which were found in the FFT search. Within the effective parameter space of our periodicity search, the FFT search recovered $\sim$98\% of the known pulsars within the survey sensitivity limits. After accounting for previously reported pulsars whose observations were not processed or whose parameters lay outside our search range, the updated GBNCC census now includes 964 pulsars. 
\end{abstract}


\section{Introduction} \label{sec:intro}
Based on the version 2.8.1\footnote{\url{https://www.atnf.csiro.au/research/pulsar/psrcat/}} of ATNF Pulsar Catalogue \citep{psrcat05a}, over 4390 radio pulsars have been discovered to date, mainly thanks to large-area radio pulsar surveys conducted with some of the world's most sensitive radio telescopes, such as the Green Bank Telescope (GBT), Parkes/Murriyang, MeerkAT, and the Five-hundred-meter Aperture Spherical radio Telescope (FAST) \citep[e.g.,][]{pmps01, keane_10, gbncc14, lotaas_19, fast_21, mmgps_23}. Such surveys have been have been responsible for the discovery of more than 90\% of the currently known radio pulsar population and remain one of the most effective ways to expand the census of neutron stars in the Galaxy. Apart from probing neutron-star formation and evolution, these surveys have uncovered a wide range of scientifically important systems. For instance, the most stable millisecond pulsars (MSPs) distributed both in the Galactic plane and at high Galactic latitudes have lead to the development of pulsar timing arrays (PTAs) for the detection of gravitational waves \citep{Nanograv+23, EPTA+2023,  Reardon+2023, CPTA+2023, Miles+2025}. The discovery of a population of relativistic binaries, intermittent and nulling pulsars, highly scattered pulsars, and steep-spectrum sources has also significantly improved our understanding of binary evolution, fundamental physics, Galactic structure, and the interstellar medium \citep[see e.g.,][]{2017ApJ...846..170T, Cromartie+2020, kramer_21, faucher_06, He+2025}. However, despite the success of these surveys, their potential is not fully realized. Pulsar searches of large-scale surveys are not only limited by their sensitivity, they are also affected by how the search is conducted. 

Pulsar searching involves several steps. The main ones include radio frequency interference (RFI) mitigation before any processing. If the data are not properly mitigated for RFI, the presence of RFI can significantly hinder real pulsar signals. Next, the data are dedispersed over a range of dispersion measure (DM) trials, and periodicity searches are performed on the dedispersed time series using either Fast-Fourier Transform \citep[FFT;][]{ransom_02} and/or Fast-Folding Algorithm \citep[FFA;][]{morello20}. If the data are not searched thoroughly enough for proper parameter space coverage, pulsars outside the search range will go unnoticed. In the post-processing stage, candidate sorting and folding must be done because pulsar searches for large-scale surveys contain tens of thousands of observations or telescope pointings, generating a very large number of candidates. Only a reduced number of candidates should pass for candidate folding. Finally, those folded candidates need to be visually inspected to identify promising pulsar candidates. 

If even one of these steps has issues, it may prevent the detection of many pulsars. Therefore, reprocessing these surveys using better search techniques is warranted. The importance of reprocessing the archival data has already been demonstrated repeatedly. For instance, the first survey that underwent reprocessing was the Parkes Multibeam Pulsar Survey (PMPS). In its first processing, 600 pulsars were found in this survey \citep{pmps01, pmps02, pmps03, pmps04}. A later reanalysis using different search algorithms identified an additional 128 pulsars, increasing the survey yield by 21\% at the time \citep{faulkner04}. Over the following two decades, the PMPS data have been reprocessed another seven times, bringing the total survey yield to 948 pulsars, a 58\% increase relative to number found in the first processing \citep{mkeith09,eatough13a, mickaliger_12, knipsel13, sengar_23}. Similarly, the reprocessing of the HTRU-S LowLat survey \citep{keith10} has increased the survey yield by 71\%, which included several different classes of pulsars \citep{sengar_22, sengar_25}. More recently, a reanalysis of archival survey data of the extragalactic satellite galaxies, the Large and Small Magellanic Clouds \citep{manchester_06, ridley_10}, resulted in the discovery of six additional pulsars (Sengar et al., accepted).

These results show that the pulsar yield of a survey can remain substantially incomplete after its first-pass processing. Since most surveys that have undergone reprocessing were conducted at 1.4~GHz, comparatively little attention has been given to reprocessing large low-frequency surveys. Reanalyzing the data is particularly important at low frequencies, where dispersion, scattering, and sky temperature generally are more severe, while the RFI environment and the strong frequency dependence of pulsar flux density present additional challenges for pulsar searches. The Green Bank North Celestial Cap (GBNCC) survey \citep{gbncc_stovall_14}, conducted at 350~MHz, therefore provides an important opportunity to investigate how much of the detectable pulsar population may remain hidden in a large low-frequency survey after its initial processing.

Motivated by this, we have reprocessed the GBNCC survey using highly effective search techniques in recovering faint, narrow-duty-cycle pulsars that might have remained undetected in the original reprocessing. GBNCC survey is well suited for this work because of its large area of the sky, low observing frequency, and a large amount of archival data. Its low-frequency sensitivity helps us find steep-spectrum and nearby pulsars, and its high-latitude coverage is useful for identifying MSPs and pulsars located far from the Galactic plane. The large number of survey pointings ($\sim 122000$) and candidates also makes GBNCC a good test case for studying how search and post-processing choices affect the final pulsar yield. This paper is organized as follows. In Section~\ref{sec:gbncc}, we briefly summarize the GBNCC survey. Section~\ref{sec:data_analysis} describes our processing pipeline, including the FFT search, candidate sifting, classification, and the direct folding search algorithm. In Section~\ref{sec:results}, we present the newly discovered pulsars and their basic properties. Section~\ref{sec:census} discusses an updated census of pulsars detected in the GBNCC survey, including the efficiency of our FFT search, a comparison with the previous GBNCC census, and redetections of pulsars subsequently found in other surveys. In section~\ref{sec:discussion}, we investigate why the new pulsars were missed in the previous processing. Finally, we summarize our conclusions in Section~\ref{sec:conclusions}.

\section{GBNCC survey} \label{sec:gbncc}

The Green Bank North Celestial Cap (GBNCC) survey is a 350~MHz pulsar survey conducted with the GBT between 2009 and 2020. The goal of this survey was to scan the full sky visible from the GBT at declinations $\delta > -40^{\circ}$, corresponding to $\sim 82$\% of the celestial sphere. This extensive sky coverage and sensitivity, makes GBNCC one of the widest and most sensitive low-frequency pulsar surveys conducted to date. Observations were taken using the Green Bank Ultimate Processing Instrument (GUPPI) at a central frequency of 350~MHz. Data were recorded in 8-bits with a 100~MHz bandwidth spanning 300--400~MHz, divided into 4096 channels, with a sampling time of 81.92$\mu$s. The survey comprised $\sim 120,000$ pointings, each with an integration time of 2-min. Its low observing frequency, makes GBNCC particularly sensitive to steep-spectrum pulsars, slow pulsars, and MSPs. Away from the Galactic plane, the survey sensitivity was estimated to be $\sim$1.1~mJy \citep{gbncc_stovall_14}. However, a later population census analysis of GBNCC found a lower limiting flux density of 0.74~mJy \citep{mcewen_19}.

The main scientific goal of this survey was to discover a large population of pulsars, particularly at high galactic latitudes, where reduced interstellar scattering and lower sky temperature provide a suitable environment for detecting MSPs. To date, GBNCC has found a diverse population of pulsars, including 27 MSPs ($P<30 \, \rm ms$), 24 rotating radio transients (RRATs), and 143 canonical/slow pulsars. Several of these pulsars are of particular astrophysical importance, including eclipsing MSPs and black-widow systems, a double neutron star system, nulling pulsars, and MSPs suitable for high-precision timing and inclusion in PTAs \citep[e.g.,][]{gbncc_stovall_14, gbncc_3_lynch_18, gbncc_6_agazie_21, gbncc_7_swiggum_23, gbncc_8_fiore_23, gbncc_9_mcewen_24}.

\section{DATA ANALYSIS AND SEARCH TECHNIQUES}
\label{sec:data_analysis}

The original GBNCC survey was conducted using 8-bit data, with a total data volume of $\sim 700$~TB. However, long-term storage of the full-resolution data became a significant challenge after the initial processing, the data were downsampled to 2-bit format, reducing the volume to $\sim 150$~TB. This data set was transferred and processed using the OzSTAR supercomputing facility at Swinburne University and the ATLAS computing cluster at the Albert Einstein Institute in Germany.

\begin{figure*}[!t]
\centering
\includegraphics[width=0.9\textwidth]{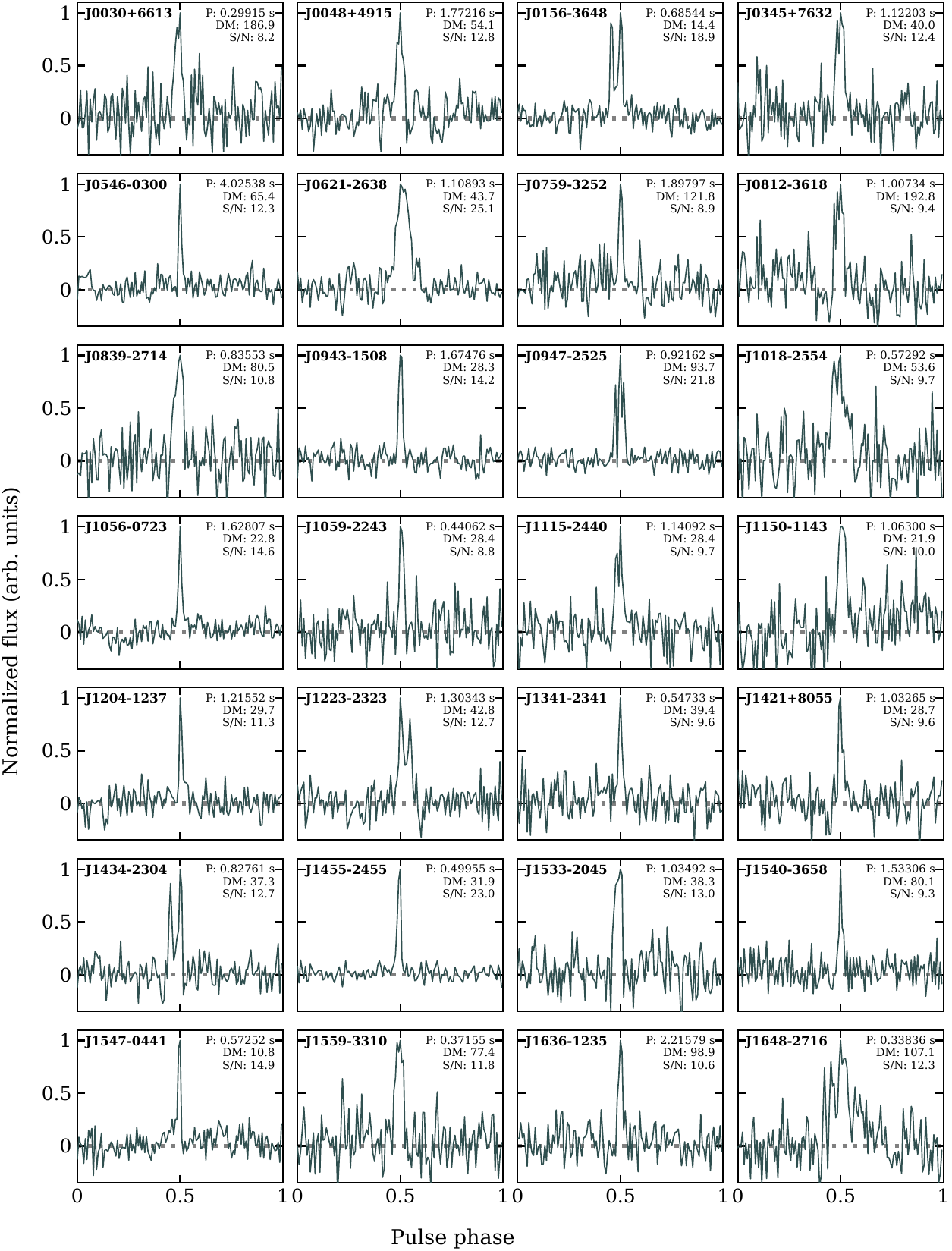}
\caption{Normalized pulse profiles of the newly confirmed GBNCC pulsars. The profiles have been centered on the brightest phase bin for display. Each panel lists the pulsar name, spin period, DM, and folded S/N.}
\label{fig:profiles1}
\end{figure*}

\begin{figure*}[!t]
\centering
\includegraphics[width=0.9\textwidth]{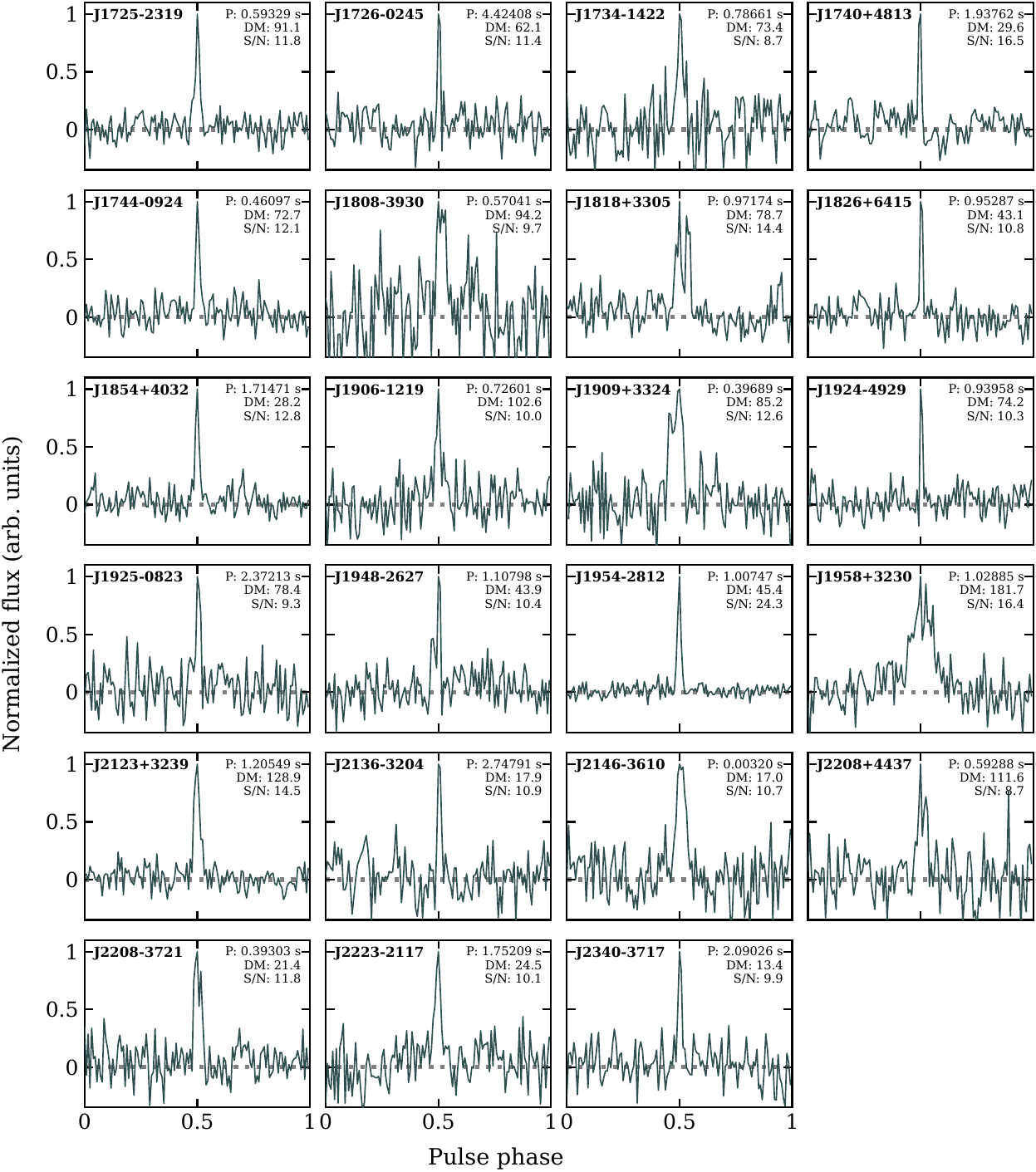}
\label{fig:profiles2}
\end{figure*}

\subsection{FFT based periodicity search}
\label{subsec:fft_search}

Before conducting the periodicity searches, we converted the \texttt{psrfits} raw data format of the GBNCC observations to \texttt{SIGPROC filterbank} format using the \texttt{digifil} tool of the \texttt{dspsr} software package. We then obtained an RFI mask for each observation  \texttt{rfifind} from the \texttt{PRESTO}\footnote{PulsaR Exploration and Search TOolkit, \url{https://github.com/scottransom/presto}}~\citep{Ransom2011} pulsar-search package. From this mask, we obtained a list of the bad frequency channels which were removed using \texttt{sigpyproc} to produce cleaned filterbank files. From the entire survey, an average of $\sim15\%$ of the frequency channels were masked in each observation. In the original processing of the survey, the 360--380~MHz frequency range was excluded before the main search because this part of the 350~MHz band was persistently effected by RFI at the GBT site. However, visual inspection of observations containing known pulsars, showed that this frequency range was not always severely affected by persistent RFI. In many observations, this part of the band was less affected by RFI and contained recoverable pulsar signals. Therefore, we did not remove this portion of the band by default in our reprocessing, and allowed \texttt{rfifind} to identify and mask contaminated channels on an observation-by-observation basis.

After the data were cleaned, we first conducted an FFT-based search using \texttt{PEASOUP}\footnote{\url{https://github.com/ewanbarr/peasoup}}, a GPU-accelerated pulsar search pipeline. We searched over a DM range of  $2$--$500~{\rm pc~cm^{-3}}$, using the \texttt{dedisp} \citep{barsdel12} library for dedispersion. This range covers the expected DMs of pulsars in the GBNCC survey because no pulsar has ever been detected with a DM below 2 $\rm pc \, cm^{-3}$ and no known pulsar in the survey has a DM above 500~$\rm pc \, cm^{-3}$. The trial DMs were generated adaptively rather than using a fixed DM step size. As a result, the spacing between adjacent trials increases with DM. The spacing was chosen such that, for a pulsar with a true DM midway between two adjacent trials, the additional pulse broadening due to the DM offset does not exceed the given tolerance. In our search, we used a DM tolerance of 1.1, corresponding to a maximum increase of 10\% in the effective pulse width and an S/N loss of only $\sim5\%$.  

\texttt{PEASOUP} can also use a time-domain resampling method to search for accelerated pulsar signals. However, since the GBNCC observations have a duration of only 120~s, we did not perform an acceleration search, as the amount of pulse smearing caused by binary acceleration is small over such a short integration. For a pulsar undergoing a constant line-of-sight (l.o.s.) acceleration $a$, the maximum number of Fourier bins over which a signal drifts during an observation is given by $z_{\rm max}=T_{\rm obs}^2a_l/(cP)$, where $T_{\rm obs}$ is the observation duration, $P$ is the pulsar spin period, and $c$ is the speed of light. Since $z$ scales as $T_{\rm obs}^2$, acceleration-induced drift is substantially reduced for the short GBNCC observations. For example, assuming a circular, edge-on binary with an orbital period of 2~hr, a double neutron star system with component masses of $1.4~{\rm M_\odot}$ has a maximum l.o.s. acceleration of $\sim300~{\rm m~s^{-2}}$. For a 20-ms pulsar, this corresponds to a drift of only $\sim0.72$ Fourier bins over a 120-s observation. Similarly, a $1.4~{\rm M_\odot}$ pulsar in a 2-hr orbit with a $0.2~{\rm M_\odot}$ companion has a maximum acceleration of $\sim62~{\rm m~s^{-2}}$, corresponding to a drift of $\sim1$ Fourier bin for a 3-ms pulsar. We therefore expect the loss of sensitivity due to binary acceleration to be small for the majority of pulsars in the GBNCC searches.


Our search summed up to 32 harmonics, and candidate lists were generated using a minimum spectral S/N of 5.0. We chose this low threshold to avoid discarding weak candidates before folding. Each observation produced, on average, $\sim10,000$ candidates, folding such a large number of candidates per observation is infeasible. Following the approach of \citet{sengar_23}, we therefore used candidate-sifting criteria based on the spin period $P$, spectral signal-to-noise ratio $\mathrm{S/N}_{\rm FFT}$, and the number of summed harmonics $n_{\rm h}$. Pulsar searches use various sifting criteria to limit the number of candidates selected for folding. Among these, a spectral S/N cut is the most common. However, $\mathrm{S/N}{\rm FFT}$ is not always a reliable indicator of the significance of a candidate, especially for faint and narrow-duty-cycle pulsars. The Fourier power of a pulsar signal is distributed across the fundamental and its harmonics, with narrow-duty-cycle pulsars having significant power extending to a larger number of higher harmonics. Although harmonic summing recovers some of this power, it is an incoherent process, and the power of the harmonic-summed fundamental signal is lower than that of its folded signal at the correct period, which represents its true S/N.

\cite{sengar_23} showed that candidates with modest $\mathrm{S/N}_{\rm FFT}$ can become considerably more significant after folding, particularly for longer-period, narrow-duty-cycle pulsars detected in multiple harmonics. The boost in S/N after folding as compared to the $\mathrm{S/N}_{\rm FFT}$ can be a factor of 1.5--2.0, or even higher. So, a faint pulsar with a folded S/N of 9.0 could have a spectral S/N$\leq$7, placing it close to the noise floor of the FFT search. Applying a single high $\mathrm{S/N}_{\rm FFT}$ threshold could therefore reject some real pulsars before they are ever folded. To avoid this, our candidate-sifting procedure considered reviewing the entire candidate list while selecting only a limited number of candidates for folding. We used progressively stricter criteria as $\mathrm{S/N}_{\rm FFT}$ decreased. Candidates found only in the fundamental, i.e. with $n_{\rm h}=0$, were selected only if  $\mathrm{S/N}_{\rm FFT}>9.5$). Relatively bright candidates with $n_{\rm h}>0$ were selected irrespective of their spin period. Candidates close to the search threshold were retained only if they occurred in period and harmonic sums ranges where pulsar-like signals are more likely. In other words, at lower $\mathrm{S/N}_{\rm FFT}$, candidates were required to have longer periods and to be detected in a larger number of harmonics. A candidate was then sifted for folding if it satisfied the following criteria:

\begin{equation}
\mathcal{C}_{\rm keep}
=
\mathcal{C}_{\rm bright}
\cup
\mathcal{C}_{\rm MSP}
\cup
\mathcal{C}_{\rm slow}.
\end{equation}

The brighter-candidate selection was defined as

\begin{equation}
\mathcal{C}_{\rm bright}
=
\left\{
\begin{array}{l}
S \geq 7.5 \wedge n_{\rm h} > 0, \\[3pt]
\vee \; S \geq 9.5 \wedge n_{\rm h} = 0 .
\end{array}
\right.
\end{equation}

The short-period candidates, the selection was

\begin{equation}
\mathcal{C}_{\rm MSP}
=
\left\{
\begin{array}{l}
0.030 \leq P \leq 0.100, \\[2pt]
7.0 \leq S < 7.5, \\[2pt]
0 < n_{\rm h} \leq 4 .
\end{array}
\right.
\end{equation}

The slow-pulsar selection was

\begin{equation}
\mathcal{C}_{\rm slow}
=
\left\{
\begin{array}{l}
P > 0.100 \wedge 7.0 \leq S < 7.5 \wedge n_{\rm h} \geq 1, \\[3pt]
\vee \; P \geq 0.200 \wedge 6.0 \leq S \leq 7.0 \wedge n_{\rm h} \geq 3, \\[3pt]
\vee \; P \geq 0.300 \wedge 5.5 \leq S \leq 6.0 \wedge n_{\rm h} \geq 3, \\[3pt]
\vee \; P \geq 1.0 \wedge 5.0 \leq S < 5.5 \wedge n_{\rm h} \geq 4 .
\end{array}
\right.
\end{equation}

Using the above candidate sifting criteria, we obtained an average of $\sim 180$ candidates per observation, a factor of $\sim 55$ fewer than in the original candidate lists. To fold these candidates, we used the similar methodology as described in \citet{sengar_23, sengar_25}. We first folded each candidate using \texttt{dspsr} to produce archive files containing the folded data as a function of time, frequency, and pulse phase. The archives files were cleaned using \texttt{clfd}\footnote{\url{https://github.com/v-morello/clfd}} \citep{morello19}, and \texttt{pdmp} from \texttt{psrchive} was then used to obtain diagnostic plots and optimize the period and DM. We discarded candidates with a folded S/N $< 8$ or optimized DM  $< 2~\rm pc \, cm^{-3}$. Candidates that survived these conditions were then refolded at zero DM using the \texttt{pam} from \texttt{psrchive}, allowing us to compare the folded S/N at the optimized DM with that at zero DM. In the absence of too much RFI, a genuine dispersed pulsar signal is expected to be stronger at its optimized DM and to weaken when dedispersed at zero DM, whereas broadband RFI often remains strong at or near zero DM. For each candidate, we recorded the original FFT search parameters, the optimized period and DM from \texttt{pdmp}, the folded S/N, pulse width, duty cycle, S/N increase relative to the FFT detection, the zero-DM folded S/N, and the ratio of the zero-DM to optimized-DM folded S/N. These quantities were used to rank candidates and aid the visual inspection of the final diagnostic plots.

\startlongtable
\label{tab:pulsar_params}
\begin{deluxetable*}{lcccccccccc}

\tabletypesize{\small}

\setlength{\tabcolsep}{4pt}
\tablecolumns{11}
\tablecaption{Parameters of the newly discovered GBNCC pulsars. The right ascension (R.A.) and declination (Decl.) are based on the GBNCC discovery pointing positions, with uncertainties set by the $\sim36\arcmin$ FWHM of the GBT beam at 350~MHz. The spin period ($P$) and dispersion measure (DM) are obtained from the search and subsequent candidate refinement. $W_{50}$ is the pulse width measured at 50\% of the profile peak. The distance, $d_{\rm NE2025}$, is estimated from the measured DM using the NE2025 Galactic free-electron density model \citep{ocker_26}. Lower limits on $d_{\rm NE2025}$ are indicated by ``$>$'' when the measured DM exceeds the maximum Galactic DM predicted along the line of sight. $N_{\rm harm}$ is the harmonic-summing index at which the pulsar was found, where $n=0,1,2,\ldots$ corresponds to summing
$2^n=1,2,4,\ldots$ harmonics. $\mathrm{S/N}_{\rm FFT}$ is the spectral S/N reported by the search pipeline, while $\mathrm{S/N}_{\rm Fold}$ is the S/N obtained after folding the candidate at its optimized period and DM.}
\tablehead{
\colhead{Jname} &
\colhead{R.A.} &
\colhead{Decl.} &
\colhead{$P$} &
\colhead{DM} &
\colhead{$W_{50}$} &
\colhead{$d_{\rm NE2025}$} &
\colhead{$S_{350}$} &
\colhead{$N_{\rm harm}$} &
\colhead{$\mathrm{S/N}_{\rm FFT}$} &
\colhead{$\mathrm{S/N}_{\rm Fold}$} \\
\colhead{} &
\colhead{(hh:mm)} &
\colhead{(dd:mm)} &
\colhead{(ms)} &
\colhead{(pc cm$^{-3}$)} &
\colhead{(ms)} &
\colhead{(kpc)} &
\colhead{(mJy)} &
\colhead{} &
\colhead{} &
\colhead{}
}
\startdata
J0030+66 & 00:30(3) & $+$66:13(18) & 299.15(1) & 186.9(3) & 13.0 & $>20.5$ & 0.9 & 4 & 5.1 & 8.2 \\
J0048$+$49 & 00:48(3) & $-$49:15(18) & 1772.16(34) & 54(1) & 76.1 & 2.7 & 0.8 & 5 & 6.8 & 12.8 \\
J0156$-$36 & 01:56(3) & $-$36:48(18) & 685.55(5) & 14.4(4) & 46.8 & 1.0 & 1.1 & 5 & 13.8 & 18.9 \\
J0345+76 & 03:45(3) & $+$76:32(18) & 1122.0(1) & 40.0(7) & 56.0 & 2.2 & 0.8 & 3 & 10.3 & 12.4 \\
J0546$-$03 & 05:46(3) & $-$03:00(18) & 4025(2) & 65(3) & 99.8 & 2.8 & 0.6 & 5 & 7.1 & 12.3 \\
J0621$-$26 & 06:21(3) & $-$26:38(18) & 1108.9(1) & 43.7(7) & 82.2 & 3.0 & 1.5 & 5 & 9.7 & 25.1 \\
J0759$-$32 & 07:59(3) & $-$32:52(18) & 1897.6(4) & 120(1) & 52.1 & 1.0 & 0.4 & 4 & 6.2 & 8.5 \\
J0812$-$36 & 08:12(3) & $-$36:18(18) & 1007.3(1) & 192.8(8) & 48.0 & 2.2 & 0.7 & 5 & 5.0 & 9.4 \\
J0839$-$27 & 08:39(3) & $-$27:14(18) & 835.53(8) & 80.5(6) & 45.5 & 0.8 & 0.7 & 3 & 6.8 & 10.8 \\
J0943$-$15 & 09:43(3) & $-$15:08(18) & 1674.8(3) & 28(1) & 45.7 & 1.9 & 0.5 & 4 & 8.2 & 14.2 \\
J0947$-$25 & 09:47(3) & $-$25:25(18) & 921.62(9) & 93.7(6) & 51.4 & $>19.6$ & 1.2 & 3 & 20.3 & 21.8 \\
J1018$-$25 & 10:18(3) & $-$25:54(18) & 572.92(10) & 53.6(10) & 51.8 & 5.1 & 0.7 & 3 & 6.3 & 9.7 \\
J1056$-$07 & 10:56(3) & $-$07:23(18) & 1628.1(3) & 23(1) & 44.9 & 1.6 & 0.6 & 5 & 9.2 & 14.6 \\
J1059$-$22 & 10:59(3) & $-$22:43(18) & 440.62(2) & 28.4(3) & 12.7 & 1.9 & 0.3 & 4 & 5.7 & 8.8 \\
J1115$-$24 & 11:15(3) & $-$24:40(18) & 1140.9(2) & 28.4(9) & 53.3 & 1.9 & 0.4 & 5 & 5.5 & 9.7 \\
J1150$-$11 & 11:50(3) & $-$11:43(18) & 1063.0(1) & 21.9(8) & 50.6 & 1.6 & 0.5 & 5 & 5.2 & 10.0 \\
J1204$-$12 & 12:04(3) & $-$12:37(18) & 1215.5(2) & 29.7(8) & 34.3 & 2.6 & 0.5 & 5 & 5.2 & 11.3 \\
J1223$-$23 & 12:23(3) & $-$23:23(18) & 1303.4(3) & 43(1) & 90.9 & 5.1 & 0.8 & 5 & 7.4 & 12.7 \\
J1341$-$23 & 13:41(3) & $-$23:41(18) & 547.33(3) & 39.4(3) & 15.9 & 3.4 & 0.5 & 5 & 7.3 & 9.6 \\
J1421+80 & 14:21(3) & $+$80:55(18) & 1032.7(1) & 28.7(6) & 30.0 & 2.3 & 0.4 & 5 & 6.2 & 9.6 \\
J1434$-$23 & 14:34(3) & $-$23:04(18) & 827.6(1) & 37.3(7) & 57.3 & 2.7 & 1.1 & 5 & 8.2 & 12.7 \\
J1455$-$24 & 14:55(3) & $-$24:55(18) & 499.55(3) & 31.9(3) & 13.6 & 2.0 & 1.3 & 3 & 15.3 & 23.1 \\
J1533$-$20 & 15:33(3) & $-$20:45(18) & 1034.9(1) & 38.3(6) & 46.6 & 2.4 & 0.8 & 5 & 9.1 & 13.0 \\
J1540$-$36 & 15:40(3) & $-$36:58(18) & 1533.1(2) & 80.1(10) & 42.5 & 5.3 & 0.8 & 4 & 5.2 & 9.3 \\
J1547$-$04 & 15:47(3) & $-$04:41(18) & 572.52(4) & 10.8(4) & 15.0 & 1.0 & 0.7 & 4 & 6.7 & 14.9 \\
J1559$-$33 & 15:59(3) & $-$33:10(18) & 371.55(2) & 77.4(2) & 17.4 & 4.1 & 1.1 & 3 & 7.3 & 11.8 \\
J1636$-$12 & 16:36(3) & $-$12:35(18) & 2215.8(5) & 99(1) & 67.4 & 3.1 & 0.7 & 4 & 7.3 & 10.6 \\
J1648$-$27 & 16:48(3) & $-$27:16(18) & 338.36(6) & 107.1(10) & 40.4 & 3.8 & 2.2 & 1 & 11.5 & 12.3 \\
J1725$-$23 & 17:25(3) & $-$23:19(18) & 593.29(4) & 91.1(4) & 18.0 & 4.2 & 1.7 & 5 & 7.0 & 11.8 \\
J1726$-$02 & 17:26(3) & $-$02:45(18) & 4424(2) & 62(3) & 107.7 & 3.4 & 0.8 & 5 & 6.1 & 11.4 \\
J1734$-$14 & 17:34(3) & $-$14:22(18) & 786.61(6) & 73.4(5) & 30.9 & 2.2 & 1.0 & 4 & 5.3 & 8.7 \\
J1740$+$48$^{\star}$ & 17:40(3) & $+$48:13(18) & 1937.6(4) & 30(1) & 47.4 & 2.3 & 0.7 & 4 & 6.1 & 16.5 \\
J1744$-$09 & 17:44(3) & $-$09:24(18) & 460.97(2) & 72.7(3) & 12.8 & 2.5 & 1.3 & 5 & 5.7 & 12.1 \\
J1808$-$39 & 18:08(3) & $-$39:30(18) & 570.41(4) & 94.2(4) & 78.9 & 4.9 & 2.4 & 5 & 6.4 & 9.7 \\
J1818$+$33 & 18:18(3) & $+$33:05(18) & 971.7(1) & 78.7(8) & 70.7 & $>20.6$ & 1.2 & 3 & 6.3 & 14.4 \\
J1826$+$64 & 18:26(3) & $+$64:15(18) & 952.87(10) & 43.1(6) & 24.1 & 4.0 & 0.5 & 3 & 8.2 & 10.8 \\
J1854$+$40 & 18:54(3) & $+$40:32(18) & 1714.7(3) & 28(1) & 48.0 & 2.2 & 0.6 & 5 & 5.5 & 12.8 \\
J1906$-$12 & 19:06(3) & $-$12:19(18) & 726.1(1) & 103.1(8) & 26.5 & 4.2 & 1.1 & 5 & 6.2 & 9.9 \\
J1909$+$33 & 19:09(3) & $+$33:24(18) & 396.89(2) & 85.2(4) & 27.8 & 6.4 & 1.2 & 4 & 6.8 & 12.6 \\
J1924$-$49 & 19:24(3) & $-$49:29(18) & 939.58(9) & 74.2(6) & 22.9 & 6.7 & 0.5 & 5 & 5.7 & 10.3 \\
J1925$-$08 & 19:25(3) & $-$08:23(18) & 2372.1(6) & 78(1) & 65.6 & 3.9 & 0.7 & 5 & 5.0 & 9.3 \\
J1948$-$26 & 19:48(3) & $-$26:27(18) & 1108.0(2) & 43.9(8) & 46.3 & 2.1 & 0.7 & 5 & 5.9 & 10.4 \\
J1954$-$28 & 19:54(3) & $-$28:12(18) & 1007.5(1) & 45.4(6) & 26.3 & 2.6 & 1.3 & 3 & 14.3 & 24.3 \\
J1958$+$32 & 19:58(3) & $+$32:30(18) & 1028.8(3) & 182(1) & 120.6 & 7.0 & 4.1 & 5 & 5.6 & 16.4 \\
J2123$+$32 & 21:23(3) & $+$32:39(18) & 1205.5(2) & 128.9(8) & 43.9 & $>24.1$ & 0.8 & 5 & 9.2 & 14.5 \\
J2136$-$32 & 21:36(3) & $-$32:04(18) & 2747.9(8) & 18(2) & 68.3 & 1.1 & 0.4 & 5 & 9.4 & 10.9 \\
J2146$-$36 & 21:46(3) & $-$36:10(18) & 3.198988(1) & 17.020(2) & 0.2 & 1.0 & 0.6 & 2 & 8.1 & 10.7 \\
J2208$-$37 & 22:08(3) & $-$37:21(18) & 393.03(2) & 21.4(3) & 16.7 & 1.4 & 0.6 & 4 & 6.8 & 11.8 \\
J2208$+$44 & 22:08(3) & $+$44:37(18) & 592.88(5) & 111.6(5) & 32.5 & 7.5 & 0.7 & 4 & 6.5 & 8.7 \\
J2223$-$21 & 22:23(3) & $-$21:17(18) & 1752.1(6) & 24(2) & 63.5 & 1.9 & 0.4 & 5 & 6.4 & 10.1 \\
J2340$-$37 & 23:40(3) & $-$37:17(18) & 2090.3(4) & 13(1) & 58.3 & 0.9 & 0.4 & 5 & 5.2 & 9.9 \\
\enddata

\tablecomments{$^{\star}$ Co-discovered in a Galactic survey for pulsars using CHIME/FRB (\url{https://catalog.chime-frb.ca/galactic}).}
\end{deluxetable*}

\subsection{Direct Folding Search Method}
\label{subsec:dfsa}

FFT-based periodicity searches are computationally efficient and form the basis of most large-scale pulsar surveys. However, their sensitivity can be reduced for long-period, narrow duty cycle pulsars, whose signals lie at the low Fourier frequencies where red noise can affect their detection \citep[see e.g.,][]{lazarus_015, van_Heerden_2016}. These effects can be mitigated by using the Fast Folding Algorithm (FFA) phase-coherent search technique, which can provide improved sensitivity to long-period pulsars \citep{morello20}.

In a typical FFA search, the raw data are first dedispersed over a set of trial DMs to produce time series. To search a range of periods, each time series is divided into segments with lengths close to the period of interest. A periodic signal whose period matches the segment length remains at the same phase, however a signal with different period shows a systematic phase drift. The FFA searches nearby periods by efficiently combining the segments with different relative phase shifts to compensate for these drift rates. By recursively reusing intermediate sums, folded profiles can be generated efficiently for many closely spaced trial periods  \citep{staelin69}, and detections with high significance are then selected across the searched period range. Detections at similar periods in neighboring DM trials are subsequently grouped into candidates.

Although FFA searches for periodic signals through phase-coherent folding, candidate detection and the generation of the final diagnostic folded data remain separate stages of the search. We therefore investigated whether direct folding itself could be used as a practical blind periodicity search technique for short observations. Rather than first identifying periodic signals using an FFT or FFA and then folding selected candidates, in our direct-folding search method (DFSM) we use a different approach in which we construct a grid of trial spin periods and DMs and directly folds the raw filterbank data at each trial combination. Each trial produces a folded archive containing the signal as a function of pulse phase, observing time, and frequency. The resulting archive is then searched locally in period and DM to determine the values that maximize the folded S/N. If a trial is sufficiently close to the true pulsar period and DM, the optimization converges toward the same period and DM, producing a significant folded detection. In this way, the periodicity search, parameter refinement, and generation of the final time--phase and frequency--phase diagnostic data are combined within the same procedure.

DFSM does not rely on a Fourier-domain detection statistic or an intermediate FFA candidate-selection stage, although it is not intrinsically immune to red noise, baseline variations, or RFI, which can still affect the folded S/N. We restricted the search to $0.5\leq P\leq10$~s and $5\leq{\rm DM}\leq200$~pc~cm$^{-3}$. The short 120-s GBNCC integrations make this restricted search computationally tractable, whereas the number of required trial folds increases rapidly for shorter periods, longer observations, and wider DM ranges. The construction of the period--DM grid is described in Appendix~\ref{appendix:dfsm}.

\begin{figure*}
\centering
  \includegraphics[width=1.95\columnwidth]{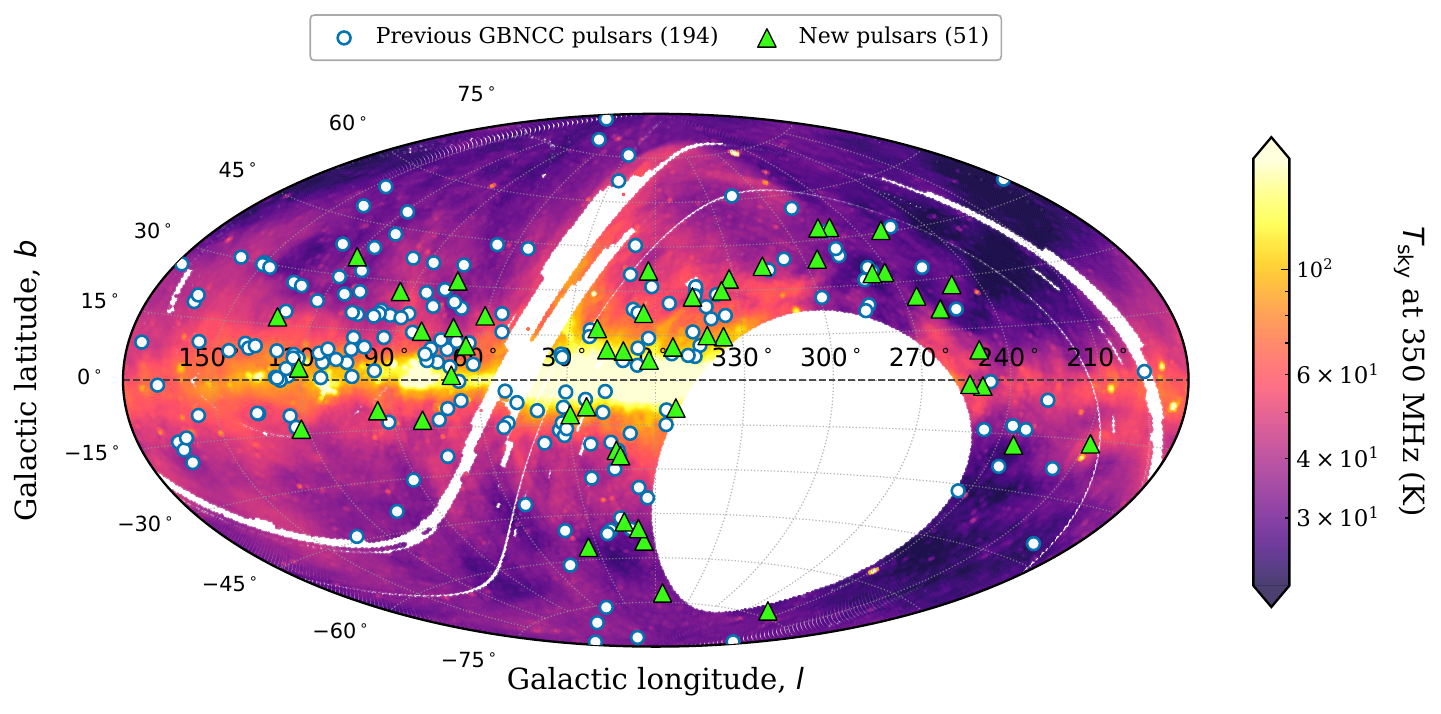}
\caption{Aitoff projection of the GBNCC survey pointings in Galactic coordinates. The background points show the GBNCC pointings, colored by the estimated sky temperature $T_{\rm sky}$ at 350~MHz, derived from the scaled Haslam 408~MHz map \citep{haslam_82}. Blue open circles mark previously reported GBNCC pulsars, while light green triangles show the 51 pulsars confirmed in our reprocessing. The newly confirmed pulsars follow the GBNCC survey footprint and are mostly located at low to intermediate Galactic latitudes, where the sky temperature, and hence the survey sensitivity varies, substantially across the sky.
}
  \label{fig:sky map}
\end{figure*}

\section{Results}
\label{sec:results}

\subsection{Discoveries using FFT search}

The entire GBNCC survey was first processed with FFT-based searches as described in Section~\ref{subsec:fft_search}. After the candidate sifting, we folded about 22 million candidates for the entire survey. Using the candidate classification procedure described above, we visually inspected $\sim 15,000$ candidates, corresponding to approximately one candidate for every eight observations. If we divide these candidates based on spin period, $\sim 3500$ candidates were visually inspected with $P>100$~ms, and the remaining $\sim$11500 were with $P<100$~ms. From our visual inspection, we found 60 promising pulsar candidates that satisfied our selection criteria. 

All but one of these candidates were slow-pulsar candidates, which was an MSP. Prior to conducting confirmation observations, we folded the nearest GBNCC pointings at the candidate periods and DMs. This allowed us to confirm 20 candidates as real pulsars, some with tentative detections and some with bright detections. To get additional confirmations, candidates with declinations $> -28~\deg$ were observed with the Canadian Hydrogen Intensity Mapping Experiment (CHIME) \citep[see e.g.,][]{CHIME/Pulsar+2021}. CHIME observed 41 candidates for an average of 15~min in the 400--800~MHz frequency band, using 1024 frequency channels and a time resolution of 390~$\mu \rm s$, sufficient for the confirmation of slow pulsars. We directly folded the CHIME data using the spin periods and DMs measured from the discovery observations. We confirmed 17 of the 22 candidates at positive declinations, while none of the candidates at negative declinations were confirmed with CHIME. The most plausible reason for this is the declination dependency of CHIME's dwell time and sensitivity. Sources at lower declinations transit farther from the zenith and have shorter effective time than sources at higher northern declinations. In addition, CHIME's effective collecting area decreases toward lower declinations. Therefore, CHIME is less sensitive to sources at low and negative declinations, making the confirmation of faint candidates more difficult.

\begin{figure}
\centering
  \includegraphics[width=0.95\columnwidth]{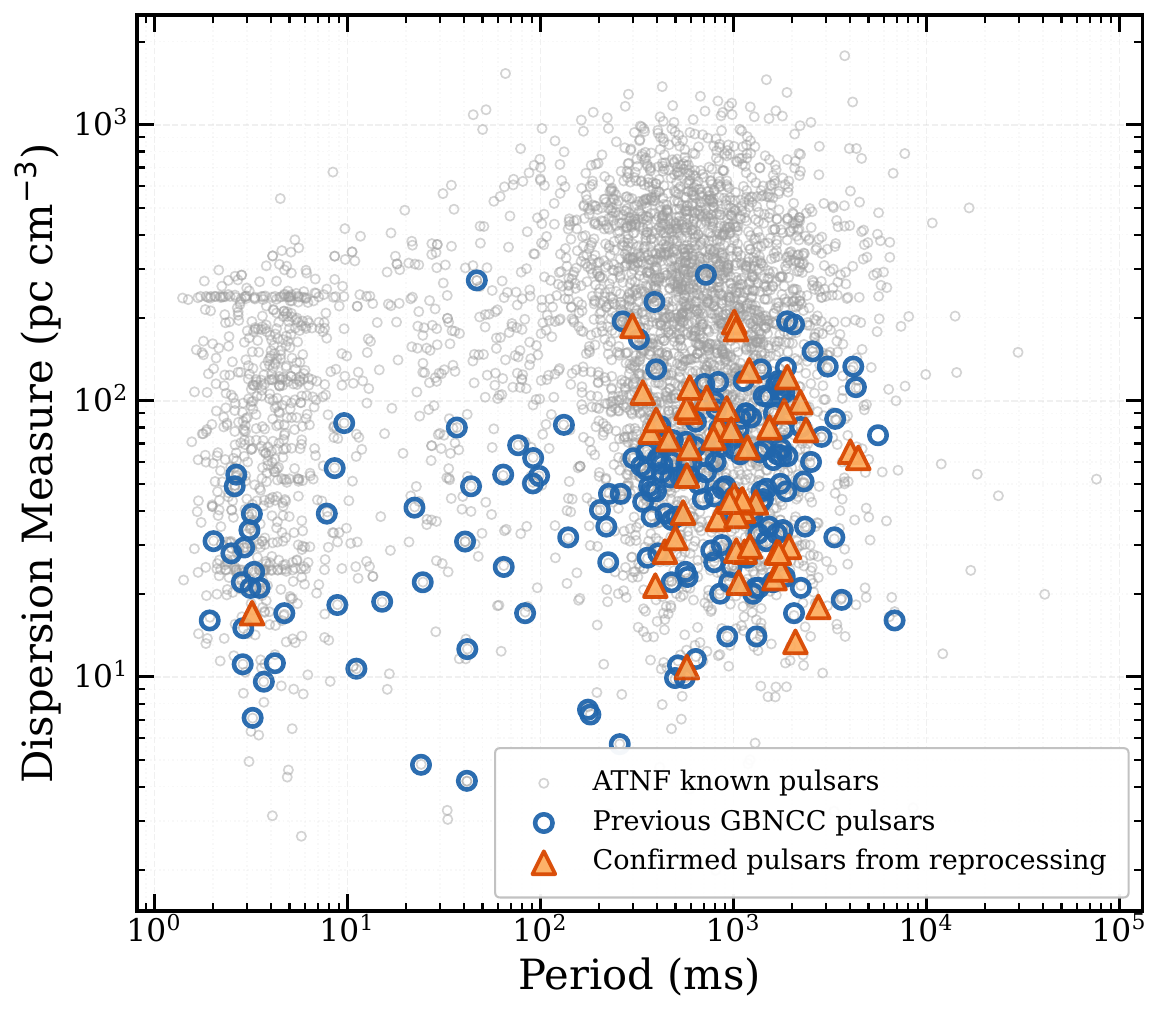}
\caption{
Period--dispersion measure distribution of the pulsars confirmed in our GBNCC reprocessing. Gray circles show known pulsars from the ATNF catalog, blue circles show previously reported GBNCC pulsars, and orange triangles mark the pulsars confirmed in this work. Both axes are shown on logarithmic scales. The newly confirmed pulsars occupy the same broad region of parameter space as previous GBNCC discoveries, with most having spin periods of a few hundred milliseconds to a few seconds and DMs between $\sim 10$ and $200~{\rm pc~cm^{-3}}$.}
  \label{fig:period dm distribution}
\end{figure} 

These 60 candidates were subsequently followed up with the GBT as part of an observing campaign to confirm them as pulsars. We obtained a total of 75 observing epochs between 2026 March and April under project GBT26A-436, using the Versatile GBT Astronomical Spectrometer (VEGAS) backend centered at a frequency of 350\,MHz (300--400\,MHz bandwidth). The data set consisted of 4096 frequency channels sampled every 82~$\mu s$ and 25~kHz channel bandwidth, using coherent dedispersion at the DM of each candidate. Integration times ranged from 2 to 8~min and were selected in order to get a minimum folded S/N value of 12 for each candidate. Of the 60 candidates observed, 51 were confirmed as pulsars. The discovery parameters of these 51 pulsars are listed in Table~\ref{tab:pulsar_params}, and their discovery pulse profiles are shown in Figure~\ref{fig:profiles1}. The sky distribution of the newly confirmed pulsars across the GBNCC survey footprint is shown in Figure~\ref{fig:sky map}.

\subsection{Discoveries using DFSM}
\label{subsec:dfsa_discoveries}

We also reprocessed the GBNCC survey using DFSM in the period range of 500~ms to 10~s and DM range of 5--200~$\rm pc \, cm^{-3}$. As described in Section~\ref{subsec:dfsa}, 3500 period--DM trials were generated for each observation, resulting in the same number of candidate plots. Since this method produced a very large number of candidates, we first discarded those with a folded S/N $ < 7.5$. The remaining candidates were then refolded at zero DM, and we calculated the ratio of the S/N at zero DM and that at the optimized DM. Candidates with a ratio below 1.0 were retained for further inspection. 

From the resulting candidates, we further selected candidates with a duty cycle $\delta_{\rm duty} < 5$\%, targeting narrow-duty-cycle signals for which FFT searches can have reduced sensitivity. We also removed observations in which a known pulsar was present. After applying these criteria, a total of $\sim$ 160\,000 candidates remained for visual inspection. 

A substantial fraction of these candidates were associated with periodic RFI detected repeatedly across the DFSM trial grid. Since DFSM folds the data independently at each trial period and DM, strong periodic RFI can be detected in several neighboring trials, which can be optimized toward nearly the same period by \texttt{pdmp}. Such signals can also be detected over a wide DM range, especially in presence of strong, narrow band RFI. Inspection of the DFSM candidates in the period--DM plane, revealed several recurrring periods extending across a broad range of DMs. These candidates were discarded, leaving a total of $\sim 15,000$ candidates for manual inspection.

During manual candidate inspection, we found all 42 pulsars that were found in our FFT searches with spin periods $ P >500$~ms. In addition, we identified PSR~J0048$+$49 with  $P=1772$~ms and $\mathrm{DM}=54~\rm pc \, cm^{-3}$. Further inspection, revealed that this pulsar was also present in our FFT results with a $\mathrm{S/N}_{\rm FFT}$ of 6.8 and a folded S/N of 12.8, but had been missed during candidate inspection.

\subsection{Pulsar properties}

Among the 51 newly discovered pulsars, 50 are slow/canonical pulsars and one is an MSP. The canonical pulsars have spin periods ranging from 0.299 to 4.424~s and DMs between 10.85 and 192.80~pc~cm$^{-3}$. They occupy a similar region of the period--DM plane to the previously discovered GBNCC pulsars (see Figure \ref{fig:period dm distribution}).  More than half of them (28/50) have spin periods longer than 1~s.

\begin{figure}
\centering
  \includegraphics[width=0.98\columnwidth]{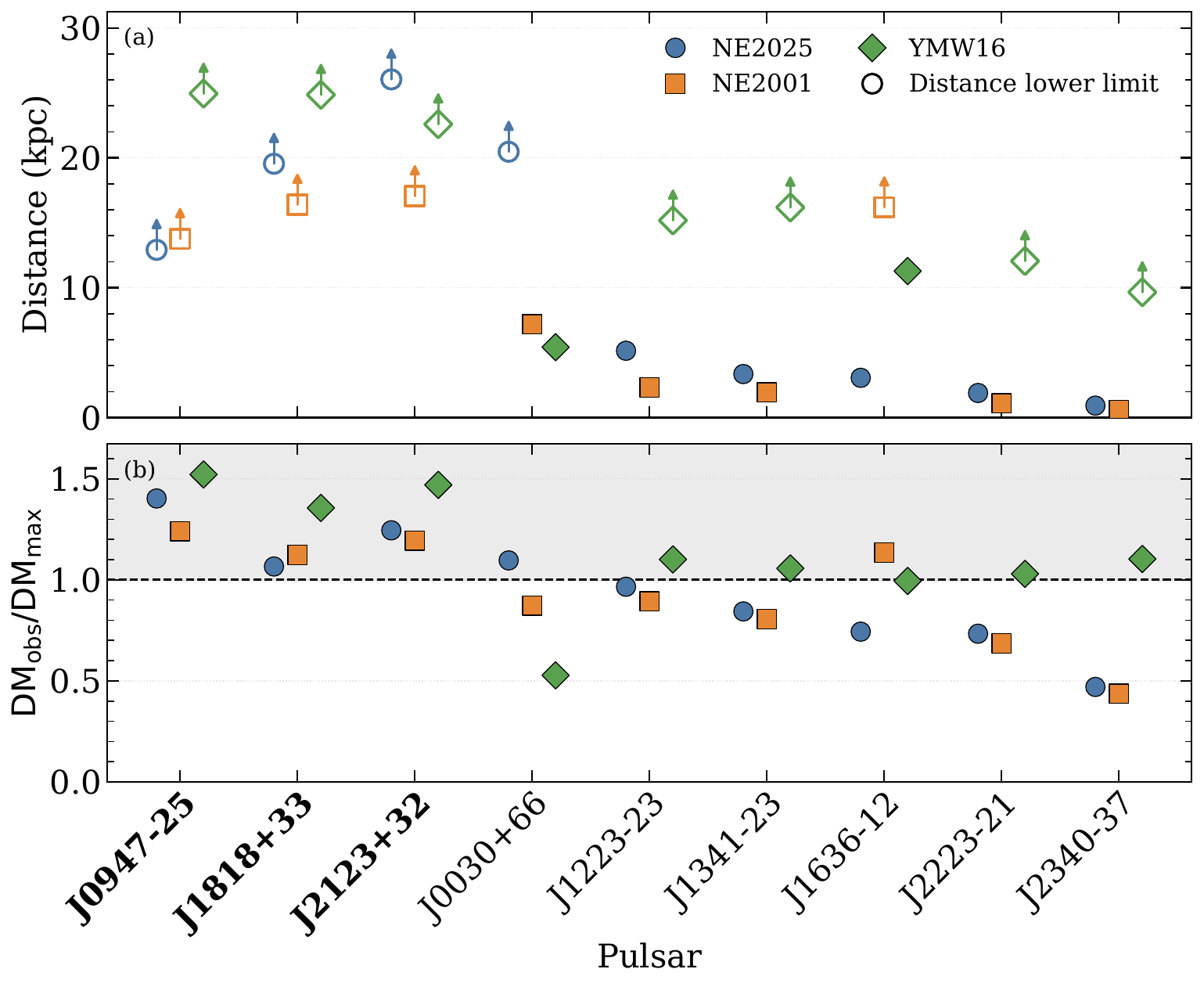}
\caption{Comparison of DM-derived distances for the nine pulsars whose observed DMs exceed the maximum Galactic l.o.s. DM predicted by at least one of the NE2025, NE2001, and YMW16 electron-density models. \textit{Top:} DM-derived distances from each model, with open symbols and upward arrows indicating lower limits. \textit{Bottom:} Ratio of the observed DM to the maximum DM predicted by each model along the l.o.s. The dashed line marks $\mathrm{DM}_{\rm obs}/\mathrm{DM}_{\max}=1$, above which the observed DM exceeds the model prediction. Bold labels identify the three pulsars for which this occurs in all three models.}
\label{fig:dm model plot}
\end{figure}

We estimated DM-derived distances using the NE2025 \citep{ocker_26}, NE2001 \citep{ne2001}, and YMW16 \citep{ymw16} Galactic free-electron density models. For the 48 pulsars with finite distance estimates, the mean (median) distances are 3.0 (2.5)~kpc, 2.2 (1.9)~kpc, and 3.5 (2.8)~kpc for NE2025, NE2001, and YMW16, respectively. However, we found that for some pulsars, the distance estimates differ substantially between models (see Figure \ref{fig:dm model plot}). In particular, the observed DMs of four pulsars exceed the maximum l.o.s. DM predicted by NE2025, four exceed that predicted by NE2001, and seven exceed that predicted by YMW16. These discrepancies are not unexpected, as the models adopt different prescriptions for the Galactic free-electron distribution and often predict different maximum DMs along a given l.o.s. \citep[e.g.,][]{price_2021}. Notably, three pulsars PSRs J0947$-$2525, J1818$+$3305, and J2123$+$3239, have observed DMs exceeding the maximum predictions in all three models (see Figure~\ref{fig:dm model plot}). This suggests that the total free-electron column density along these l.o.s may be underestimated, possibly due to ionized structures that are not adequately represented in the models \citep[e.g.,][]{ocker_26}.

\begin{figure*}
\centering
  \includegraphics[width=2.1\columnwidth]{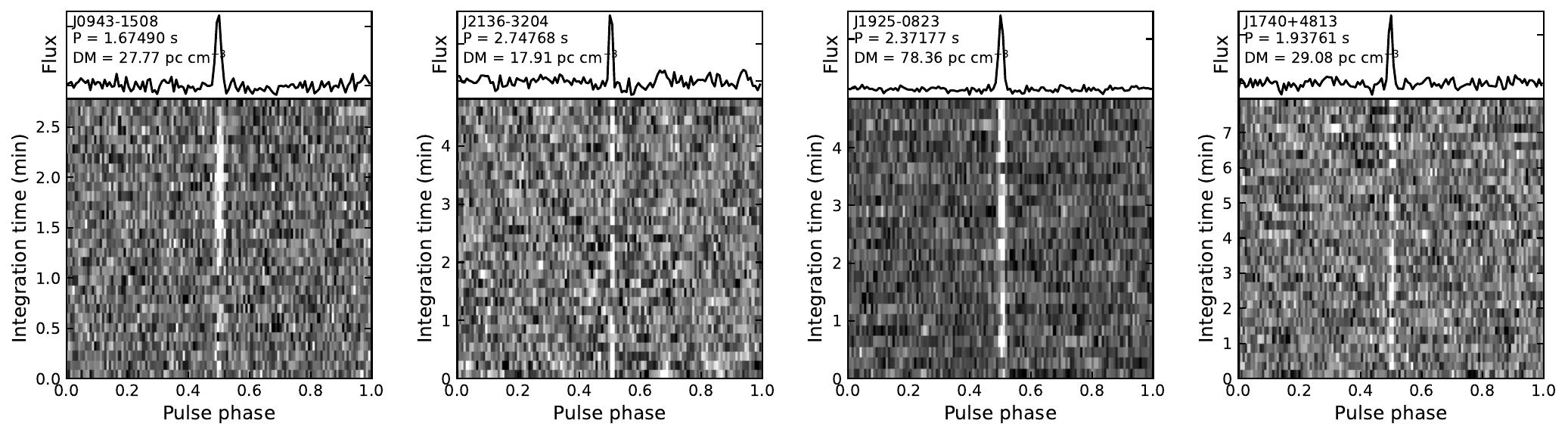}
\caption{
Confirmation observations of four newly discovered pulsars showing clear evidence for nulling. For each pulsar the upper panel shows the integrated pulse profile, while the lower panel shows the pulse intensity as a function of pulse phase and integration time.}
\label{fig:nulling psrs plots}
\end{figure*} 

We also find evidence of nulling in many of these sources. From visual inspection of the discovery and confirmation observations, 19 of the 50 canonical pulsars show either clear or tentative evidence for nulling. Examples of four of the brighter sources showing clear nulling in their
confirmation observations are shown in Figure \ref{fig:nulling psrs plots}. Since the available observations are relatively short in duration, we did not measure nulling fractions, as longer observations are required to characterize their emission behavior. About 15 pulsars also show double or multiple-component pulse profiles (see Figure~\ref{fig:profiles1}).

Using the discovery pulse profiles, we estimated the mean flux density of each pulsar using the radiometer equation \citep{1985ApJ...294L..25D},

\begin{equation}
S_{350} =
\frac{\beta\,({\rm S/N})\,(T_{\rm sky}+T_{\rm rec})}
     {G\sqrt{n_{\rm p}\,t_{\rm int}\,\Delta \nu}}
\left(\frac{W_{50}}{P-W_{50}}\right)^{1/2},
\end{equation}

\noindent where $S_{350}$ is the estimated flux density at 350~MHz, $G=2.0~{\rm K~Jy^{-1}}$ is the GBT gain, $n_{\rm p}=2$ is the number of summed polarizations, $\Delta\nu=80~{\rm MHz}$ is the effective observing bandwidth, and $\beta=1.2$ accounts for digitization and other instrumental losses. We adopted a receuver noise temperature of $T_{\rm rec}=23$~K and estimated $T_{\rm sky}$ at each pulsar position from the 408~MHz all-sky map scaled to 350~MHz. $W_{50}$ values were measured from Gaussian fits to the discovery pulse profiles, and the corresponding duty cycle was taken to be $W_{50}/P$. 

\begin{figure}
\centering
  \includegraphics[width=0.99\columnwidth]{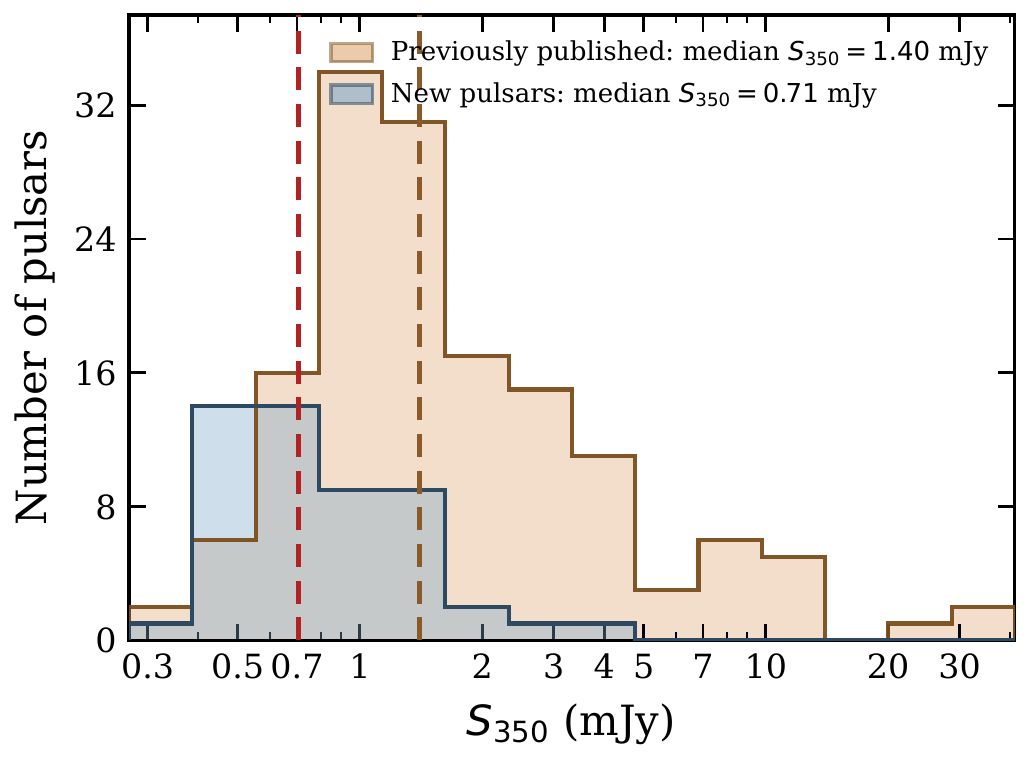}
\caption{Distribution of $S_{350}$ for the 51 newly confirmed pulsars and previously published GBNCC pulsars with available cataloged 350-MHz flux densities. The dashed vertical lines mark the median values of $0.71$~mJy and $1.40$~mJy for the new and previously published pulsars, respectively.}
  \label{fig:flux density hist}
\end{figure}

The resulting $S_{350}$ distributions are shown in Figure~\ref{fig:flux density hist}. The newly confirmed pulsars have a median estimated $S_{350}$ of $0.7~{\rm mJy}$. For comparison, we also show previously published GBNCC pulsars with available cataloged $S_{350}$ measurements, which have a median  of $1.4~{\rm mJy}$. This suggests that the newly confirmed pulsars are generally fainter than the previously known population. The $S_{350}$ measurements for both populations may be affected by  nulling, scintillation, intrinsic variability, and residual RFI. For the newly confirmed pulsars, an additional uncertainty arises from their poorly constrained positions within the $36'$ GBT beam. Since precise timing positions are not yet available, we cannot determine their offsets from the beam center or correct for the corresponding reduction in telescope gain. Therefore, their $S_{350}$ values should be regarded as approximate discovery flux densities. 

The only MSP discovered in our reprocessing is PSR~J2147$-$3604, with a spin period of $\sim$3.2~ms and a DM of $\sim17$~pc~cm$^{-3}$. It is also the only pulsar among the new discoveries for which we have obtained a phase-connected timing solution. Dedicated timing observations spanning a year show that this pulsar is in a $\sim4.5$-day binary orbit. The pulsar shows strong interstellar scintillation and has also been detected in gamma rays. Its radio profile is narrow, with a duty cycle of $\sim 6.5$\%. The short spin period and narrow profile make PSR~J2147$-$3604 a promising candidate for high-precision timing and a potential addition to PTAs, although its long-term timing stability still needs to be established. A detailed timing study of this system will be presented in Vleeschower et al.\ (2026, in preparation).

Another interesting source is PSR~J1954$-$28, which we detected in the blind periodicity search with $P=1.00747$~s and DM$=45.4$~pc~cm$^{-3}$. A previously reported GBNCC RRAT, PSR~J1956$-$28, was discovered in the original single-pulse processing of the survey, and with a reported $P=0.2600144$~s and DM$=45.7$~pc~cm$^{-3}$ \citep{karako_argaman_15}. However, the reported positions of these two pulsars are separated by $\sim21$~arcmin which is within the GBT beam size of 36~arcmin at 350~MHz. We also conducted a single pulse search for PSR~J1954$-$28 and detected individual pulses from this source. The similar DMs and overlapping positional uncertainties  suggest that PSR~J1954$-$2812 and PSR~J1956$-$28 may be the same source. 

For RRATs detected through only a small number of individual pulses, the rotation period is generally estimated from common divisors of the time separations between the detected pulses. With only a few pulses, the inferred period can be ambiguous and can be revised when additional pulses or periodic emission are detected \citep{karako_argaman_15,keane_11}. Revised periods that are not simple harmonics of the initially reported value have previously been found for some sparsely detected RRATs \citep[e.g.][]{keane_11}. Therefore, PSR~J1954$-$28 is most likely a redetection of PSR~J1956$-$28, with a previously incorrectly determined rotation period. A timing position for PSR~J1954$-$2812 will be required to establish this association unambiguously.

\section{GBNCC pulsar census}
\label{sec:census}

To evaluate of the completeness of the survey and the effectiveness of the reprocessing, we analyzed the known pulsar population in the GBNCC survey. The previous census by \citet{mcewen_19} used  observations obtained up to late 2018, containing 108\,000 pointings. Of the 1413 pulsars known present in the GBNCC survey region at the time, 670 were detected (including 66 MSPs), corresponding to a recovery fraction of $\sim47\%$ of the pulsars known at that time. This analysis was primarily based on folding observations using the known pulsar periods and DMs. However, the fraction of these pulsars that would have been recovered independently by the FFT-based periodicity or single-pulse search pipelines was not quantified.

We therefore repeated the known-pulsar census using the complete GBNCC data set of $\sim$122\,000 pointings covering $\sim36\,430$~deg$^2$, and the currently known pulsar population listed in the ATNF Pulsar Catalogue. We searched for known pulsars using two different methods. First, we cross-matched the known pulsar population with the candidates found from our blind FFT-based periodicity search. Second, we performed an independent direct-folding search. For each GBNCC pointing, we identified all pulsars in the catalog located within $1\deg$ of the pointing center and directly folded the corresponding observation at the known spin period and DM of each pulsar. This approach allowed us to identify known pulsars present in the GBNCC observations that were not found in our FFT-based search.  Table~\ref{tab:census_comparison} summarizes the results of both censuses, noting that the known-pulsar population and available survey data have increased since the previous analysis.

\begin{table*}
\centering
\caption{Comparison of the previous GBNCC pulsar census with the present analysis. The known-pulsar  populations are not identical because the pulsar catalog and the available GBNCC data set have both increased since the previous census.}
\label{tab:census_comparison}
\begin{tabular}{lcc}
\hline
\hline
Quantity & \citet{mcewen_19} & This work \\
\hline
GBNCC pointings available
    & $\sim108,000$ & $\sim122,000$ \\
Fraction of completed survey
    & $\sim83\%$ & $\sim100\%$ \\
Sky coverage
    & $\sim30,237$ deg$^2$ & $\sim36,430$ deg$^2$ \\
Known pulsars in survey region
    & 1413 & 2553\\ 
Known-pulsar detections/redetections
    & 670 & 855 \\
MSPs among known-pulsar detections
    & 66 & 97 \\
New pulsars reported in this work
    & -- & 51 \\
Total GBNCC pulsars
    & 670 & 964 \\
Pulsars per pointing
    & 1 per 161 & 1 per 127 \\
Surface density
    & 1 per 45.1 deg$^2$ & 1 per 37.8 deg$^2$ \\
\hline
\end{tabular}

\end{table*}

\subsection{FFT search efficiency}
\label{sec:census_recovery}


Excluding 51 pulsars reported in this work, our FFT-based search recovered 826 known pulsars in 1977 individual pointings, including 168 previously reported GBNCC pulsars. Although, previously 194 pulsars have been found by the GBNCC survey (27 MSPs, 24 RRATs, and 143 normal pulsars). The remaining 26 previously reported GBNCC pulsars were not recovered in our FFT search. Eighteen of these are RRATs, while the observations containing the other eight pulsars were not processed in the present reprocessing due to processing error.


By directly folding the observations using the known pulsar periods and DMs, we redetected a total of 855 known pulsars. Of these, 826 were also recovered by our blind FFT-based search, while the remaining 29 were not present in the FFT candidate lists. Of these 29 pulsars, five were in observations that had not been processed through our pipeline due to processing errors, four have spin periods longer than 5~s and therefore fall outside the period range of our analysis, and three are RRATs, for which detection in a periodicity search is not always guaranteed. The remaining 17 pulsars were within the searched parameter space but were not recovered by our FFT pipeline. Half of these non-detections have low folded S/N values of ($7 < \mathrm{S/N} < 9$), while most of the remaining sources have ($9 < \mathrm{S/N} < 12$), with the exception of two pulsars that have folded S/N values of $\sim 25$. After excluding the 12 pulsars outside the effective search parameter space, 826 of the remaining 843 known pulsars were detected in our FFT search, corresponding to a recovery fraction of approximately 98\%. Therefore, only $\sim 2$\% of the known pulsars were missed in our FFT search.

Some of the pulsars that were only found through direct folding are interesting because of their extremely long spin periods. In particular PSR~J0311+1402, with $P\simeq40.9$~s and DM $\simeq 19\,\rm pc\,cm^{-3}$. This pulsar was discovered serendipitously by \citet{wang_25} in a 2-min ASKAP/CRACO observation originally carried out for testing interplanetary scintillation measurements. Only three consecutive pulses, each with a duration of $\sim$0.5~s were detected, from which a spin period of  $\sim$40.9~s was obtained. Follow-up observations showed that the source is highly intermittent. It was subsequently detected with MeerKAT, but only one of five Parkes observations resulted in a detection, while a one-hour ASKAP observation produced only a weak folded detection. More recent observations at 111~MHz also show that the emission can be extremely sporadic, with only a few individual pulses detected over several thousand observing sessions and no periodic detection in an FFA search \citep{tyulbashev_25}. At low frequencies, the source therefore appears to exhibit RRAT-like behavior or a very high nulling fraction. In the GBNCC data, we found this pulsar in two pointings separated by 12~min, with a folded S/N of 270. 

The 855 known pulsars found in our analysis include a larger MSP population than in the previous GBNCC census. We redetected 89 previously known MSPs, and with the addition of the new MSP discovered in this reprocessing, the total number of MSPs detected in the survey data increases to 90.

\subsection{Comparison with the previous GBNCC census}

Comparing our results with the previous GBNCC census reported by \citet{mcewen_19}, we found that 638 of the 670 pulsars are present among the 855 pulsars detected in our analysis. For the 32 pulsars reported as detections by \citet{mcewen_19} but not detected in our analysis, we found no detections in either the FFT or directly folding the data at known period and DM. Seven of these sources are MSPs and the remaining are slow pulsars. To investigate these non-detections, we further refolded the nearest available GBNCC pointing to the position of each pulsar reported in the ATNF Pulsar Catalogue. However, we did not find any convincing detections for any of these pulsars, despite 10 reported to have S/N $>15$ in \citet{mcewen_19}. One plausible explanation is that the closest available observation is not necessarily the same pointing in which the pulsar was detected by \citet{mcewen_19}. For 15 of the 32 pulsars, the angular separation between the pulsar position and the center of the nearest available pointing exceeds 18~arcmin. It is therefore possible that these pulsars were detected in different, and closer GBNCC pointings that were included in the earlier census but are not present in the data set available for our reprocessing. Although conversion of the data from 8-bit to  2-bit introduces a modest $\sim10\%$ loss in sensitivity loss, but it doesn't explain why these pulsars were not detected as earlier census was also conducted using 2-bit conversion.

\subsection{Updated census, archival detections, and survey yield}
\label{sec:census_updated}

Combining 51 new pulsars reported in this work, 26 previous GBNCC pulsars (18 RRATs and 8 slow pulsars), and the 32 known pulsars which were previously reported in the GBNCC census, and whose relevant observations were likely not available in our reprocessing, the total pulsar census of the GBNCC survey stands at
\[
855 + 26+ 32 + 51 = 964
\, \rm pulsars\]

Of these 964 pulsars, we found 189 known pulsars that were not reported in the previous GBNCC census (see Table \ref{tab:known_gbncc_pulsars}). This difference is partly due to almost full dataset used in our analysis, since the previous census was based on GBNCC observations available through late 2018. For 60 of these 189 pulsars, the corresponding GBNCC observations were conducted after 2018 and were therefore not part of the dataset analyzed by data set analyzed by \citet{mcewen_19}. Apart from this, 59 sources are of particular interest, including seven MSPs, for which the GBNCC observations were obtained several years before their subsequent discovery in other surveys. The GBNCC observations precede the reported discovery years of these pulsars by 3--15~yr, with a median gap of 9~yr. Therefore, these pulsars were already present in the GBNCC survey data well before they were eventually discovered elsewhere. A large fraction of these pulsars was discovered in FAST surveys, including 16 in CRAFTS survey, two in GPPS \citep{gpps_2021}, two in HGLPS \citep{hglps_2024}, and one in observations of the globular cluster M15 \citep{m15_2024}. Others were found in LOTAAS \citep{lotaas_19}, CHIME \citep{good_21}, and PUMPS surveys\citep{pumps_22}, respectively. These detections demonstrate that a considerable population of detectable pulsars remained unidentified in the GBNCC data.

\begin{figure*}
\centering
  \includegraphics[width=1.3\columnwidth]{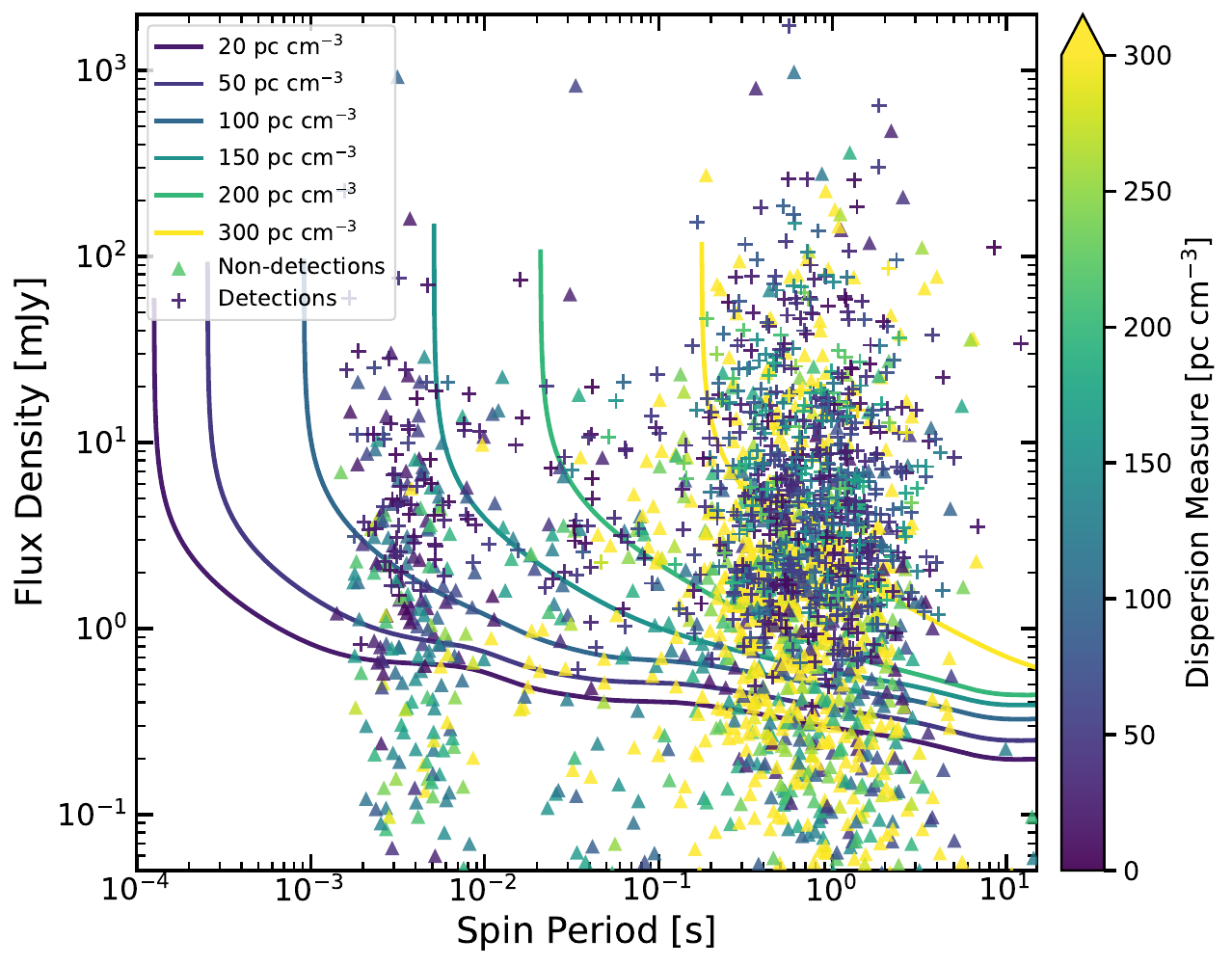}
\caption{Estimated 350~MHz flux density, $S_{350}$, as a function of spin period for all known pulsars in the GBNCC survey region for which a $S_{350}$ was obtained. Plus symbols show pulsars detected in the GBNCC data, and triangles show non-detections. Points are color-coded by pulsar DM. The solid curves show representative GBNCC sensitivity limits for DMs of 20, 50, 100, 150, 200, and 300~pc~cm$^{-3}$, calculated using an empirical period-dependent duty-cycle relation derived from the running median of cataloged pulse widths. These curves represent nominal survey sensitivity and are not strict detection boundaries for individual pulsars, since the effective sensitivity also depends on local sky temperature, usable bandwidth, beam offset, scintillation, intrinsic variability, and uncertainties in the estimated flux densities of the non-detections.}
  \label{fig:gbncc_dm_curves}
\end{figure*}

The increase in the number of pulsar detections in the GBNCC survey is also apparent when  normalized by the number of survey pointings. The data set used by \citet{mcewen_19} contained approximately 108\,000 pointings and 670 detected pulsars, corresponding to one pulsar per 161 pointings. In comparison, the present census of 964 pulsars from  $\sim 122,000$ pointings corresponds to one pulsar per 127 pointings. The number of pulsars per pointing is therefore $\sim27\%$ higher than in the previous census. 

A similar comparison can be made in terms of sky coverage. The completed GBNCC survey covers $\sim 36\,430$~deg$^2$. Since the data used by \citet{mcewen_19} represented approximately 83\% of the completed survey, the corresponding sky coverage was $\sim 30\,237$~deg$^2$, their 670 pulsars give a surface density of one pulsar per 45.1~deg$^2$. The present census of 964 pulsars over the completed survey gives one pulsar per 37.8~deg$^2$, or $\sim0.026$ pulsars~deg$^{-2}$. The pulsar surface density obtained in the updated census is therefore $\sim19.0$\% higher than that inferred from the earlier GBNCC census.

Finally, we compared the GBNCC census with the current known-pulsar population within the survey footprint. Using a matching criterion of 30~arcmin around the GBNCC pointings, we identified 2553 known pulsars within the surveyed region (excluding globular cluster pulsars). Given the census of 964 pulsars in the GBNCC data, this corresponds to a detection fraction of 39.4\%.

\begin{figure}
\centering
  \includegraphics[width=0.98\columnwidth]{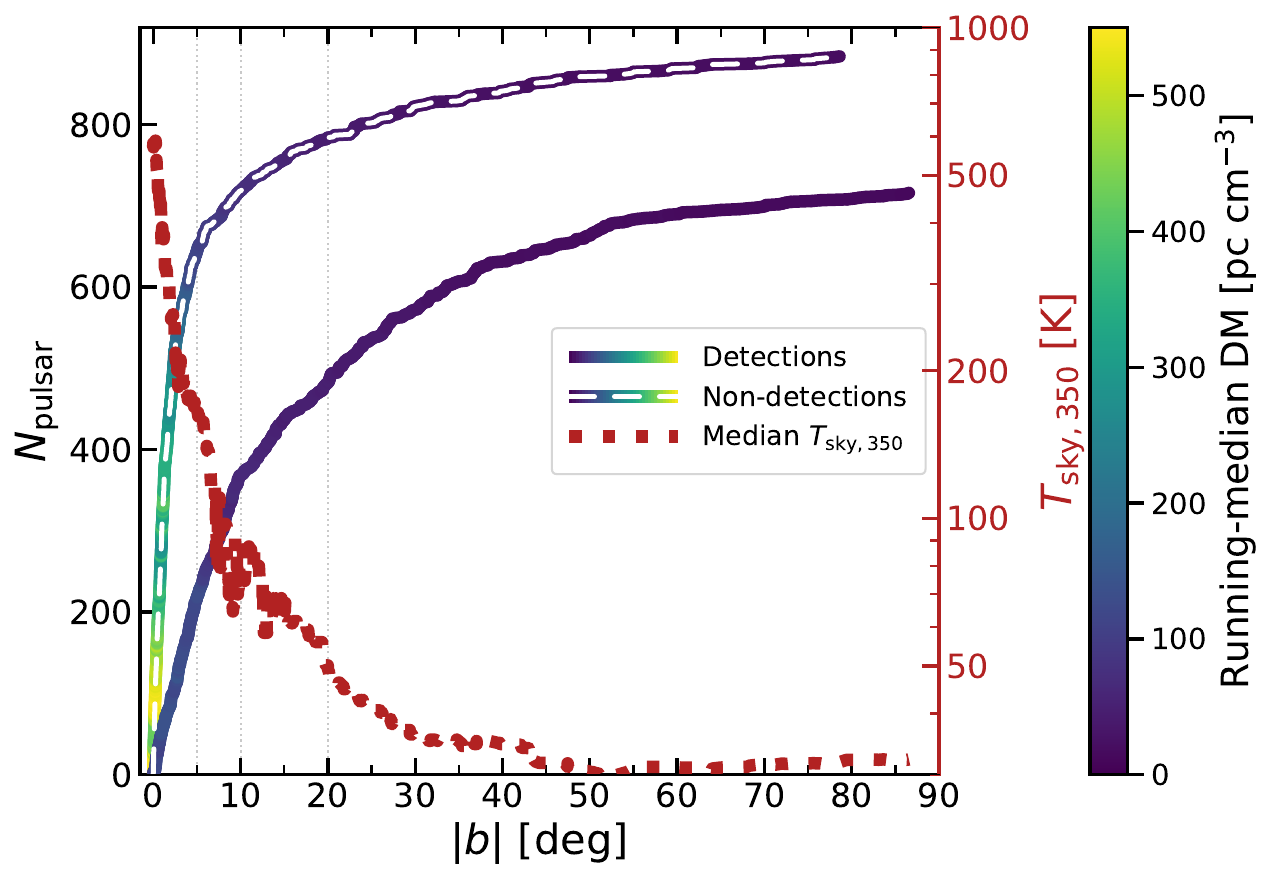}
\caption{
Cumulative number of detected and non-detected pulsars in the GBNCC as a function of absolute Galactic latitude, $|b|$. Both cumulative curves are color coded by the running median DM of their respective populations, and the common color scale shown at right. The curve corresponding to detections is shown as a continuous color-coded line, while the non-detection curve dashed white overlay on the same DM color scale. The red dotted curve shows the running-median sky temperature at 350 MHz, $T_{\rm sky,350}$, for the combined population, with values given by the right-hand axis.} 
  \label{fig:gbncc_gb_dm_sky}
\end{figure}

\subsection{Survey Sensitivity}
\label{subsec:_sensitivity}

To investigate whether detected and undetected pulsars in the GBNCC survey region occupy a distinct region of parameter space, we compared both populations in spin period, DM, sky temperature and estimated $S_{350}$.  For known pulsars detected in the GBNCC data, we estimated $S_{350}$ using the radiometer equation as described in Section 4.3, with an additional correction for attenuation within the GBT beam. We assumed a circular Gaussian beam response with a FWHM of 30 arcmin and corrected the measured flux density according to the angular separation between the catalogued pulsar position and the GBNCC pointing center.

For the non-detected pulsars, $S_{350}$ was estimated from available catalogued radio flux densities and spectral information. Where a catalogued $S_{350}$ measurement was available, we used it directly. Otherwise, if a catalogued spectral index, $\alpha$, was available, we scaled the available flux-density measurement closest in frequency to 350~MHz assuming a power-law spectrum, $S_{\nu}\propto\nu^{\alpha}$. If no catalogued $\alpha$ was available but both $S_{400}$ and $S_{1400}$ were reported, we derived $\alpha$ from these two measurements. For pulsars with at least one usable flux-density measurement but no spectral-index information, we assumed $\alpha=-1.4$. Pulsars for which no usable catalogued radio flux-density measurement was available were excluded from comparisons involving $S_{350}$.

In Figure~\ref{fig:gbncc_dm_curves} we show the estimated $S_{350}$ as a function of spin period where curves show GBNCC sensitivity limits at DMs of 20, 50, 100, 150, 200, and 300~pc~cm$^{-3}$ which we obtained using pulse width as a function of spin period using the running median of the catalogued $W_{50}/P$ distribution, rather than using a fixed duty cycle for all period ranges. In Figure~\ref{fig:gbncc_dm_curves}, it can be seen that the non-detected pulsars in the GBNCC survey systematically have higher DMs than the detected population. The detected pulsars have a median DM of $\simeq56.8$~pc~cm$^{-3}$, whereas the non-detected population has a median DM of $\simeq202$~pc~cm$^{-3}$, a factor of $\sim3.6$ higher. Based on KS test the difference between these populations is also significant with $p\ll0.001$.

This difference is also clearly visible in the Galactic distribution of the two populations. In Figure~\ref{fig:gbncc_gb_dm_sky} we show the cumulative number of detected and non-detected pulsars as a function of absolute Galactic latitude, $|b|$. The non-detected population is more strongly concentrated toward low Galactic latitudes i.e. $\sim$ 85\% of the non detected pulsars have  $|b|<15 \, \rm deg$ where the running-median DM and the sky temperature at 350 MHz are also largest. The median $T_{\rm sky,350}$ is $\simeq65$ K for the detected pulsars and $\simeq257$ K for the non-detected population. Both DM and $T_{\rm sky}$ decrease toward higher Galactic latitudes, where the observing conditions for a low-frequency pulsar survey become progressively more favorable. The non detections at higher $|b|$, where both DM and $T_{\rm sky,350}$ are lower are most likely due to scintillation, intermittency or nulling, and RFI in general. 

\section{Discussion}
\label{sec:discussion}

Since the majority of the new pulsars reported in this work are relatively faint, we investigated whether they occupy a distinct region of the search parameter space compared with known pulsars detected in the GBNCC observations. To make the two samples more comparable, we restricted the analysis to pulsars with spin periods $P>50$~ms and folded S/N in the range $8<{\rm S/N}_{\rm fold}<15$. No brighter pulsars were included as they can cause bias in this analysis.

\begin{figure}
\centering
  \includegraphics[width=0.98\columnwidth]{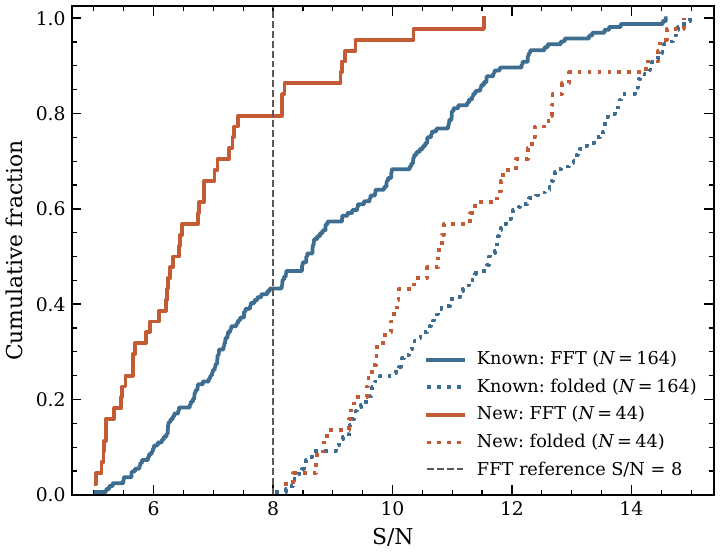}
\caption{Cumulative distributions of the Fourier-domain and folded S/N for previously known pulsars recovered in the GBNCC data (blue) and pulsars newly identified in this reprocessing (orange). The comparison is restricted to sources with $P>50~\mathrm{ms}$ and $8 \leq (S/N)_{\mathrm{fold}} < 15$. Solid lines show $S/N_{\mathrm{FFT}}$, while dotted lines show $(S/N)_{\mathrm{fold}}$. The vertical dashed line marks the reference threshold $S/N_{\mathrm{FFT}}=8$. Although the two samples have similar folded-S/N distributions, the newly identified pulsars are concentrated toward lower $S/N_{\mathrm{FFT}}$, with many lying near or below the reference threshold.}
\label{fig:sn_cumulative}
\end{figure}

Figure~\ref{fig:sn_cumulative} shows the cumulative distribution of spectral and folded S/N for the two samples. The main difference is reflected in the spectral/FFT ${\rm S/N}$, with the new pulsars located toward the lower values and many lying close to the lower end of the periodicity search sensitivity. Using a two-sample Kolmogorov--Smirnov test confirms that the spectral/FFT ${\rm S/N}$ distributions of the new and known pulsars differ significantly, with a $p$-value of $2.5\times10^{-6}$. In contrast, the folded S/N distributions show no statistically significant difference between the two samples ($p$-value=0.156).

This comparison suggests that the new pulsars were not missed because they are too faint in the GBNCC observations, but because many were detected with low S/N in the Fourier-domain. By including candidates deeper into the low-$\rm S/N_{\rm FFT}$ regime and folding these candidates, our reprocessing recovered them at higher S/N after folding.
The increased pulsar yield therefore arises in part from more permissive candidate selection and more effective use of post-search information, rather than from any improvement in the sensitivity of the observations.

\begin{figure}
  \includegraphics[width=0.95\columnwidth]{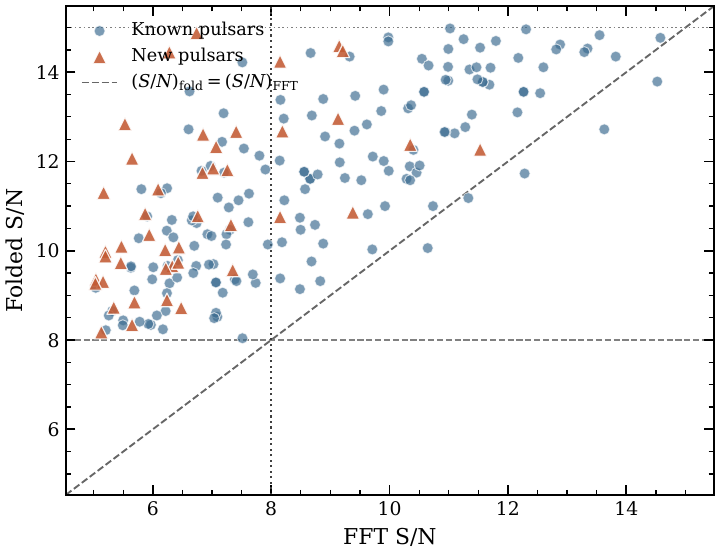}
    \includegraphics[width=0.95\columnwidth]{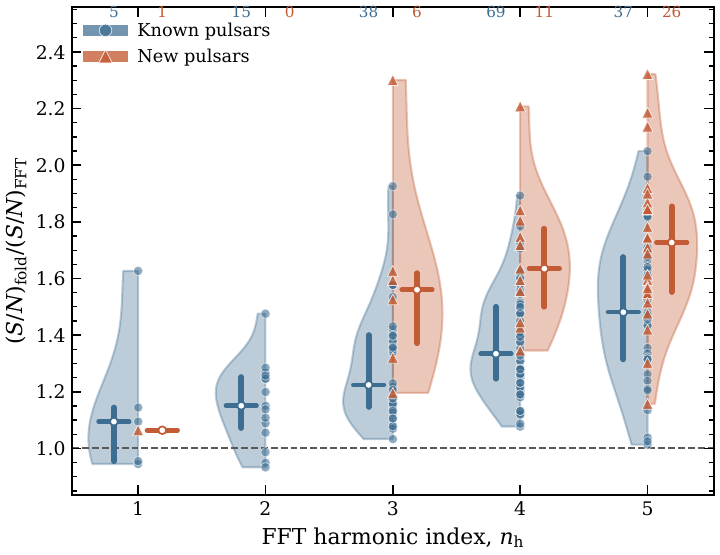}
\caption{Comparison of the spectral and folded S/N for previously known GBNCC pulsars (blue circles) and new pulsars identified in this reprocessing (orange triangles). The samples are restricted to $P>50~\mathrm{ms}$ and $8 \leq (S/N)_{\mathrm{fold}} < 15$. Top: Folded S/N as a function of $S/N_{\mathrm{FFT}}$. The dashed line represents $(S/N)_{\mathrm{fold}}=S/N_{\mathrm{FFT}}$. Most sources lie above this relation. Bottom: Folding gain, defined as $(S/N)_{\mathrm{fold}}/S/N_{\mathrm{FFT}}$, as a function of the harmonic-summing index $n_{\mathrm{h}}$. Here, $n_{\mathrm{h}}=0$ denotes detection at the fundamental, while $n_{\mathrm{h}}=1,\ldots,5$ correspond to sums of $2,\ldots,32$ harmonics. The violin profiles show the distributions within individual harmonic groups, the points represent individual pulsars, and the horizontal bars mark the medians. Pulsars detected in higher harmonic sums generally exhibit larger folding gains.}
\label{fig:sn_nh_plots}
\end{figure}

This difference is also evident when we compare $\rm S/N_{\rm FFT}$ with $\rm S/N_{\rm fold}$ (see top panel of Figure \ref{fig:sn_nh_plots}). The majority of the pulsars become substantially more significant after folding, such that a source that appears only marginally in the FFT search can become a clear pulsar detection once folded. A similar effect was reported by \citet{sengar_23, sengar_25} in the reprocessing of large scale surveys conducted at Parkes, where candidates with $\rm S/N_{\rm FFT}\sim5-6$, particularly those detected in higher harmonic sums, produced much stronger detections after folding. These results show that genuine pulsars can lie close to the FFT noise floor and may therefore be missed if candidate selection relies mainly on spectral S/N.

The folding gain, $(S/N)_{\rm fold}/(S/N)_{\rm FFT}$, also tends to be larger for candidates recovered at higher harmonic indices (see the bottom panel of Figure~\ref{fig:sn_nh_plots}). This trend can be understood in terms of pulse widths. Narrow pulses distribute Fourier power across several harmonics, therefore, a large number of harmonics are needed to maximize the detection significance. However, incoherent harmonic summing combines Fourier power without preserving the relative phase information, while each additional harmonic also contributes noise, and progressively higher harmonics contain decreasing amounts of signal power. Therefore, folding the same candidate at the correct period and DM instead combines the pulse signal coherently in phase, resulting in a stronger detection. This explains why pulsars detected in the 32-harmonic sum show, on average, a larger folding gain than those detected using fewer harmonics.

These results are particularly important for modern pulsar surveys, in which data are being recorded at very high resolution, resulting in a very large number of candidates and often require a higher $\rm S/N_{\rm FFT}$ cutoff before folding. For example, the TRAPUM surveys of the Small and Large Magellanic Clouds folded candidates with $\rm S/N_{\rm FFT}\geq9$ to reduce the number of candidates requiring folding and improve processing speed \citep{carli_24, prayag_26}. Similarly, the MPIfR--MeerKAT Galactic Plane Survey used a cutoff of $\rm S/N_{\rm FFT}=8.5$ \citep{mmgps_23}. 

Such thresholds provide a compromise between computational cost and search sensitivity, but they can also introduce selection effects that may potentially result in incomplete pulsar census, bias inferred luminosity distributions, as well as missing scientifically interesting systems. In our reprocessing, candidate selection closer to the noise floor for folding, together with additional candidate selection based on period and harmonic summing, significantly helped us to discover new pulsars that could have been rejected by a single spectral S/N threshold.

\section{Conclusion}
\label{sec:conclusions}

In this work, we have reprocessed the GBNCC survey using sensitive FFT-based periodicity searches, candidate selection and folding, and an independent direct-folding search method. Using these techniques, we obtained 60 compelling candidates and confirmed 51 pulsars with their detection in either neighboring GBNCC pointings, or new observations using CHIME or with the GBT followup observations. This represents a confirmation success rate of $\sim83\%$. Previous searches using similar techniques have also achieved candidate confirmation rates of $>80\%$ for slow pulsars \citep{sengar_23, sengar_25}. When compared with the 143 slow pulsars previously discovered in the GBNCC survey, these new pulsars represent a substantial $\sim35\%$ increment. Many of the new discoveries show evidence for nulling or strong variability. We also found a new MSP, PSR~J2147$-$3604 which is a radio-gamma ray binary pulsar. The newly discovered pulsars are predominantly faint, but otherwise occupy the same broad period--DM region as the previously known GBNCC population.

We have also conducted a new census of the GBNCC survey and accounting for pulsar detections in previous GBNCC census work \citep{mcewen_19}, there is a total of 964 pulsars, including 97 MSPs in the survey. Of these, our analysis suggests that another 59 pulsars, including seven MSPs could have been discovered in the first pass processing of the survey if the search techniques employed in this work had been used at that time. This suggests that the GBNCC survey could have discovered an additional 110 new pulsars, rather than the 51 reported in this work. The newly confirmed pulsars have a median estimated flux density of $S_{350}\simeq0.7$~mJy, whereas the previously discovered GBNCC pulsars have a median $S_{350}\simeq1.4$~mJy. This difference does not necessarily imply that the newly discovered pulsars are intrinsically fainter, because precise timing positions are not yet available for them and their discovery flux densities cannot be corrected for attenuation caused by unknown offsets within the GBT beam.

Repeated reprocessing of large-scale pulsar surveys has shown that a substantial number of pulsars (especially narrow duty cycle pulsars) can remain unidentified in the original analyses, with the GBNCC survey being the latest example. Although in any large scale survey, missing a small fraction of pulsars is expected due to limitations in search algorithms and candidate inspection. Our results indicate that an additional source of incompleteness is the way candidates from FFT-based searches are selected for folding and subsequently judged as astrophysical signals. In particular, the new pulsars have lower spectral S/N than previously known pulsars, and similar folded-S/N distributions. This shows that faint detectable pulsars can be lost at the candidate-selection stage if spectral S/N is used too aggressively as a criterion for folding. In our reprocessing, the combination of summing up to 32 harmonics and selecting candidates close to the Fourier noise floor based on their spin period and harmonic sums allowed many of these sources to be recovered.

This interpretation is supported by the pulsar census of the GBNCC presented in this work. We compared detections from our blind FFT search with those obtained by direct folding at the known period and DM. The FFT search recovered 97.8\% of the known pulsars detected through direct folding, with remaining non-detections potentially affected by strong red noise. This suggests that, when combined with candidate selection extending toward the Fourier noise floor, the FFT search is already highly efficient over the period range occupied by most canonical pulsars. The main limitation is therefore not necessarily the Fourier search itself, but the loss of faint candidates before they are folded.

An FFA based analysis would provide an important complementary check of the completeness of the GBNCC slow-pulsar population. The phase-coherent nature of the FFA has been proven to improve sensitivity to long-period and narrow duty cycle pulsars relative to FFT searches which suffer red noise and signal detection is based on incoherent harmonic summing. However, the FFA itself is not immune to red noise, and baseline variations. Therefore, it is unclear how much additional sensitivity an FFA search would provide for GBNCC. The $\sim$98\% detection rate of pulsars by our FFT search provides a useful benchmark for such a comparison. An FFA may recover some of the missing sources that were absent from the FFT candidate lists, but this is not guaranteed. Apart from this, detections of those missing pulsars alone would not establish that the FFA is more complete, since it would also be necessary to determine whether the same FFA search recovers the 97.8\% already detected by the FFT pipeline. A source-by-source comparison using the full known-pulsar sample would therefore provide a direct measurement of the relative completeness of the two approaches and reveal how their performance depends on spin period, duty cycle, and the noise properties of the observations.

Our DFSM results provide an initial, although not equivalent, indication of what a folding-based search might add. In our DFSM searches, all new 42 pulsars found by the FFT search within the period range 500~ms to 5~s were detected, while the only additional pulsar found was shown to be present in the FFT candidate list and had been missed during visual inspection of FFT search candidates. This suggests that, over much of the canonical-pulsar parameter space explored here, a sufficiently deep FFT search combined with inclusive candidate selection can already recover most detectable periodic signals. However, a full reprocessing of the GBNCC survey with an FFA would nevertheless be valuable, both to determine whether a small population remains genuinely inaccessible to the FFT search and to establish, using the same data set, the relative strengths and limitations of the two methods.

\begin{acknowledgments}
R.S thanks the continuing support from the Max Planck Society. L.V., D.K., and S.M.R. are thankful for the continuing support from the NSF Physics Frontiers Center award numbers 1430284 and 2020265. The Green Bank Observatory is a facility of the National Science Foundation (NSF) operated under cooperative agreement by Associated Universities, Inc.
S.M.R. and I.H.S. are CIFAR Fellows.  Pulsar research at UBC is supported by an NSERC Discovery Grant. 

The Green Bank Observatory is a facility of the National Science Foundation operated under cooperative agreement by Associated Universities, Inc.
The National Radio Astronomy Observatory (NRAO) is a facility of the National Science Foundation operated under cooperative agreement by Associated Universities, Inc.

We acknowledge that CHIME is located on the traditional, ancestral, and unceded territory of the Syilx/Okanagan people. We are grateful to the staff of the Dominion Radio Astrophysical Observatory, which is operated by the National Research Council of Canada. CHIME operations are funded by a grant from the NSERC Alliance Program and by support from McGill University, University of British Columbia, and University of Toronto. CHIME was funded by a grant from the Canada Foundation for Innovation (CFI) 2012 Leading Edge Fund (Project 31170) and by contributions from the provinces of British Columbia, Québec and Ontario. The CHIME/FRB Project, which enabled development in common with the CHIME/Pulsar instrument, was funded by a grant from the CFI 2015 Innovation Fund (Project 33213) and by contributions from the provinces of British Columbia and Québec, and by the Dunlap Institute for Astronomy and Astrophysics at the University of Toronto. Additional support was provided by the Canadian Institute for Advanced Research (CIFAR), the Trottier Space Institute at McGill University, and the University of British Columbia. The CHIME/Pulsar instrument hardware is funded by the Natural Sciences and Engineering Research Council (NSERC) Research Tools and Instruments (RTI-1) grant EQPEQ 458893-2014. 

\end{acknowledgments}



%
\facilities{Green Bank Telescope}

\software{astropy \citep{astropy_13,astropy_18}, PSRCHIVE \citep{psrchive04,psrchive12}, Pulsar Survey Scraper \citep{Kaplan2022}, PyGEDM \citep{price_2021}}

\appendix

\section{Implementation of the direct-folding search method}
\label{appendix:dfsm}


The DFSM introduced in Section~\ref{subsec:dfsa} was designed to find slow pulsars in the GBNCC data by directly folding the filterbank data at trial periods an DMs. The simple idea behind DFSM is that if the data are folded at an arbitrary period and DM to produce time-versus-phase and frequency-versus-phase data cubes or folded archive files with a limited number of subintergrations, frequency channels and phase bins, then these can then be searched locally over a range of neighboring period and DM range to obtain an optimized value of period and DM.

We generate the trial-period grid by limiting the maximum amount of pulse phase drift allowed across the full observation. Let $T_{\rm obs}$ be the observing time, and let $P_n$ be the current central trial period in the grid. Therefore, the number of turns/pulses accumulated over the observation is $T_{\rm obs}/P$. Therefore, if the data are folded with an offset period, $P_{\rm off}$, the accumulated phase difference or phase drift $\Delta \phi$ between the two periods in time $T_{\rm obs}$ is:
\begin{equation}
\Delta \phi =
\frac{T_{\rm obs}}{P_n}
-
\frac{T_{\rm obs}}{P_{\rm off}}
=
\left(
\frac{1}{P_n}
-
\frac{1}{P_{\rm off}}
\right)
T_{\rm obs}.
\end{equation}


Solving the above expression for $P_{\rm off}$ gives
\begin{equation}
P_{\rm off} =
\left[
\frac{1}{P_n}
-
\frac{\Delta \phi}{T_{\rm obs}}
\right]^{-1}.
\end{equation}

The difference between $P_{\rm off}$ and $P_n$ then
defines the half-width of the period interval covered by this trial,

\begin{equation}
\delta P_n =
\left| P_{\rm off} - P_n \right| .
\end{equation}

Thus, the trial period $P_n$ is taken to represent the range
\begin{equation}
P_n - \delta P_n
\leq
P
\leq
P_n + \delta P_n .
\end{equation}
The upper edge of this interval is
\begin{equation}
P_n^{+} = P_n + \delta P_n .
\end{equation}

\begin{figure}
\centering
  \includegraphics[width=0.5\columnwidth]{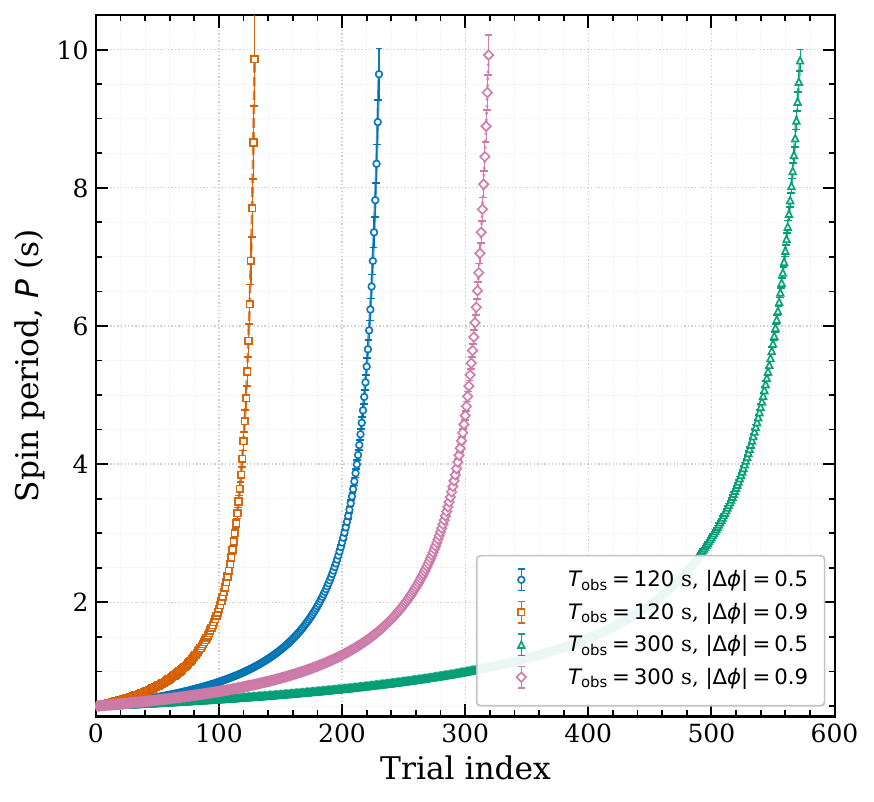}
\caption{
Period trial grids used in the DFSM. The points show the central trial periods, and the vertical error bars show the period interval covered by each trial. Curves are shown for $T_{\rm obs}=120$ and 300~s and allowed phase drifts of $|\Delta\phi|=0.5$ and 0.9 turns. Increasing $T_{\rm obs}$ makes the period grid denser, while increasing $|\Delta\phi|$ allows a coarser grid. The period search range is $0.5 \leq P \leq 10$~s.
}
  \label{fig:period_trials}
\end{figure}

This same value is then used as the lower edge of the interval represented by the next trial period. In other words, the next central trial period, $P_{n+1}$, is chosen so that a period equal to $P_n^{+}$ would drift by the allowed amount $\Delta\phi$ relative to $P_{n+1}$ over the observation. This ensures that adjacent period intervals touch each other without leaving a gap. Using the same signed phase-drift convention, the next central trial period $P_{n+1}$ is
\begin{equation}
P_{n+1} =
\left[
\frac{1}{P_n^{+}}
+
\frac{\Delta \phi}{T_{\rm obs}}
\right]^{-1}.
\end{equation}

This procedure is repeated until the trial period reaches the maximum search period, $P_{\max}$. In our search, the period grid extended from $P_{\min}=0.5$~s to $P_{\max}=10$~s. For the 120~s GBNCC observations, we used $\Delta\phi=-0.9$, corresponding to an allowed accumulated phase drift of $0.9$ turns over the duration of the observation. We used $\Delta\phi=-0.9$ to  keep the phase drift below one complete rotation. If $|\Delta\phi|\gtrsim1$, the pulse would drift through an entire phase cycle relative to the trial period, causing substantial smearing in the initial fold and introducing a phase wrap that can make subsequent period optimization less robust. The sign of $\Delta\phi$ determines the direction of the offset period calculation, while the size of the covered period interval is given by $\left|P_{\rm off}-P_n\right|$.

Coresponding to each trial spin period, we determined the DM spacing by estimating the amount of residual dispersive smearing that could be tolerated before the folded sensitivity fell below a specified threshold. The intrinsic pulse width was approximated using a fixed duty cycle,
\begin{equation}
W_{\rm int} = \delta_{\rm duty} P ,
\end{equation}
with $\delta_{\rm duty}=0.027$ which is the average of the known slow pulsars with $P>$100 ms. For a DM offset $\Delta {\rm DM}$, the effective pulse width is estimated as
\begin{equation}
W_{\rm eff} =
\left[
W_{\rm int}^{2}
+
\left(
8.3 \times 10^{6}
\Delta {\rm DM}
\frac{\Delta \nu}{\nu^{3}}
\right)^{2}
\right]^{1/2},
\end{equation}
where $\Delta \nu$ is the observing bandwidth and $\nu$ is the observing frequency. The relative folded sensitivity then is estimated as 
\begin{equation}
\eta \propto
\left(
\frac{P-W_{\rm eff}}{W_{\rm eff}}
\right)^{1/2},
\end{equation}

\begin{figure}
\centering
  \includegraphics[width=0.5\columnwidth]{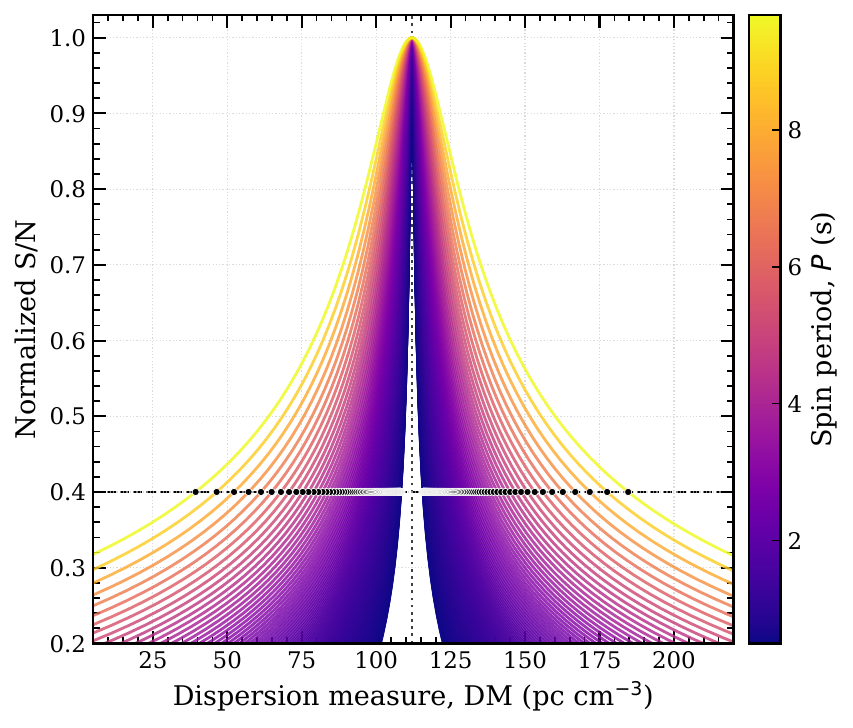}
\caption{
Normalized S/N as a function of trial DM for different trial spin periods. The curves show the expected loss in folded S/N as the trial DM moves away from the reference value, here taken to be ${\rm DM}=112~{\rm pc~cm^{-3}}$ and marked by the vertical dotted line. The color scale denotes the spin period, with broader curves corresponding to longer-period pulsars, for which a given amount of dispersive smearing produces a smaller fractional loss in S/N. The horizontal dashed line marks the adopted fractional S/N threshold of 0.4, and the black points indicate the corresponding DM offsets at which each period curve crosses this threshold. These offsets were used to define the DM trial spacing adopted for candidate refolding and optimization.}
  \label{fig:dm_curves}
\end{figure}

with $\eta$ set to zero when $W_{\rm eff} > P$. The DM step for each trial period was chosen as the point at which the normalized sensitivity fell below a specified threshold. We used a constant threshold of 0.4 for all trial periods, which means a pulsar signal would lose 60\% of it's sensitivity if folded at a DM offset corresponding to the edge of the trial-DM interval. We chose this to keep the number of DM trials manageable while still retaining a detectable signal in the initial fold. Keeping the DM interval from becoming too broad was also important for the
DM optimization, since strong RFI can sometimes produce a larger folded S/N than a nearby weak pulsar and cause the
optimization to converge away from the pulsar DM.

In Figure~\ref{fig:dm_curves} we show DM offsets for different trial periods at which the normalized S/N falls to the adopted threshold. Since $W_{\rm eff}$ increases with period, longer-period pulsars can tolerate a larger amount of residual DM smearing. Therefore, the DM spacing increases with longer trial periods, hence short-period trials require closely spaced DM values to avoid excessive pulse broadening, whereas long-period trials can be searched with coarser DM spacing while maintaining the same allowed fractional sensitivity loss. We used a DM range from 5-200 $\rm pc \, cm^{-3}$ and for each trial period, there were different trial DMs. In total, for a period range of 0.5~s to 10~s, we obtained 114 trials period and 3500 total trials.

Each observation was folded using 3500 period--DM trials. Since the procedure required only direct folding, the raw filterbank data were substantially reduced beforehand: the 4096 frequency channels were downsampled to 128 channels, and the time resolution was reduced to 512~$\mu \rm s$. This decreased data volume by a factor of 72, resulting in filterbank files of only 22~MB and significantly reducing the computational cost of the folding process. For each observation, 3500 folded archive files were generated using \texttt{dspsr}. The resulting folded archives were then searched over the period and DM offsets around the trial values using \texttt{pdmp} to obtain the optimized period and DM, and to produce diagnostic plots.

Although this technique appears to be effective, as demonstrated in Section\ref{subsec:dfsa_discoveries}, there are a few caveats. First, the number of trial periods increases rapidly when searching for shorter spin periods or processing longer observations. For a fixed allowed phase drift, the period half-width covered by a single trial is approximately

\[
\delta P \simeq \frac{|\Delta\phi|P^2}{T_{\rm obs}} .
\]

Therefore, the required period spacing becomes increasingly small toward shorter spin periods and longer integration times, resulting in a much denser trial-period grid. Figure~\ref{fig:period_trials} shows how the trial-period grid changes with $T_{\rm obs}$ and $|\Delta\phi|$.

At the same time, the required DM spacing becomes smaller at shorter trial periods. Assuming a fixed $\delta_{\rm duty}$, the intrinsic pulse width $W_{\rm int}$, decreases with decreasing spin period. Therefore, even a small DM offset can introduce dispersive smearing that is large relative to the intrinsic pulse width. To keep the fractional broadening small, a finer DM grid is therefore required at shorter periods. These considerations make this method most practical for longer periods as well as for regions of the sky where the maximum l.o.s DM is relatively low, such as at high Galactic latitudes.

\startlongtable
\begin{deluxetable*}{lcccccccc}
\tabletypesize{\small}
\tablecaption{Previously known 189 pulsars detected in the reprocessing of the GBNCC survey that were not included in previous GBNCC census.\label{tab:known_gbncc_pulsars}}
\tablewidth{0pt}
\tablehead{
\colhead{PSR} & \colhead{$P$} & \colhead{DM} & \colhead{$\theta$} & \colhead{MJD} & \colhead{$W_{50}$} & \colhead{S/N} & \colhead{$S_{350}$} & \colhead{Survey} \\
\colhead{} & \colhead{(s)} & \colhead{(pc cm$^{-3}$)} & \colhead{(arcmin)} & \colhead{} & \colhead{(ms)} & \colhead{} & \colhead{(mJy)} & \colhead{}
}
\startdata
J0011+08 & 2.55287 & 24.9 & 6.2 & 58705 & 79.7 & 8.7 & 0.4 & 1,2 \\
J0023+0923 & 0.00305 & 14.33 & 0.9 & 55325 & 0.1 & 50.1 & 2.5 & 3,4,2 \\
J0036-1033 & 0.90001 & 23.1 & 21.4 & 57061 & 28.1 & 16.4 & 1.8 & 5 \\
J0109+11 & 0.4327 & 17 & 16.3 & 58384 & 20.3 & 9.4 & 0.8 & 6 \\
J0115+6325 & 0.52146 & 65.07 & 17.6 & 55198 & 24.4 & 7.5 & 1.5 & 7 \\
J0152+0948 & 2.74665 & 22.88 & 15 & 58530 & 21.5 & 18.4 & 0.6 & 8,1 \\
J0154+1833 & 0.00236 & 19.8 & 16.2 & 58776 & 0.1 & 27.2 & 2.5 & 1,3 \\
J0158+21 & 0.5053 & 20 & 16.2 & 58769 & 23.7 & 13.5 & 1.3 & 1,7 \\
J0203-0150 & 0.00517 & 19.22 & 8.4 & 57656 & 1.1 & 19.9 & 2.7 & 9 \\
J0206-4028 & 0.63055 & 13 & 37.5 & 57035 & 14.8 & 11.4 & 7 & 10,11,12 \\
J0229+20 & 0.8069 & 27 & 13.4 & 58770 & 25.2 & 10.4 & 0.7 & 1,2 \\
J0302+2252 & 1.20716 & 18.99 & 11.8 & 58742 & 56.6 & 74 & 6 & 11,1,7,2 \\
J0304+1932 & 1.38758 & 15.66 & 8.3 & 58833 & 65 & 186.1 & 12.6 & 11,1,13,7,10 \\
J0311+1402 & 40.9107 & 19.9 & 7.6 & 58254 & 319.3 & 273.9 & 6.5 & 11 \\
J0340+4130 & 0.0033 & 49.58 & 0.6 & 55255 & 0.3 & 42.9 & 4.4 & 3,4,2 \\
J0408+551 & 1.837 & 55 & 35.6 & 55328 & 903.2 & 25.5 & 140.3 & 4,7 \\
J0435+2749 & 0.32628 & 53.19 & 9.4 & 58621 & 7.6 & 17.7 & 0.9 & 8,1,7 \\
J0452-3418 & 1.66512 & 19.78 & 13.4 & 56960 & 52 & 11.3 & 0.6 & 5 \\
J0457+23 & 0.5049 & 59 & 13.6 & 58774 & 51.3 & 10.1 & 1.4 & 1,7,2 \\
J0509+0856 & 0.00406 & 38.33 & 13 & 58594 & 0.9 & 26.8 & 5.2 & 1 \\
J0517+2212 & 0.22237 & 18.7 & 14.9 & 58778 & 33 & 68.7 & 13.8 & 8,7,2 \\
J0525+1115 & 0.35444 & 79.41 & 11.7 & 58469 & 16.6 & 142.2 & 12.6 & 10,1,13,7,11 \\
J0528+2200 & 3.74554 & 50.87 & 12.9 & 58804 & 263.4 & 240.3 & 31.4 & 14,1,15,13,7 \\
J0529-0715 & 0.68922 & 87.3 & 13.5 & 57387 & 10.8 & 22.1 & 1 & 9 \\
J0538+2817 & 0.14316 & 39.57 & 16.3 & 58739 & 10.1 & 46.3 & 7.9 & 8,1,13,7,2 \\
J0540+4542 & 0.40114 & 56.12 & 4.5 & 55553 & 59.5 & 23.3 & 3 & 9 \\
J0543+2329 & 0.24597 & 77.7 & 15.9 & 58775 & 5.8 & 103.3 & 9.9 & 16,8,14,13,7,2 \\
J0546+2441 & 2.84385 & 73.81 & 15.7 & 58776 & 44.4 & 35 & 2.5 & 8,1,7 \\
J0553+4111 & 0.55949 & 37.9 & 10.5 & 55631 & 26.2 & 13.1 & 1 & 9 \\
J0609+2130 & 0.0557 & 38.73 & 15.9 & 58831 & 2.6 & 24 & 3.3 & 8,7 \\
J0611+30 & 1.41209 & 45.26 & 17.6 & 58621 & 66.2 & 37.5 & 4.5 & 11,1,7 \\
J0614+2229 & 0.33496 & 96.91 & 3.8 & 58779 & 10.5 & 291.7 & 23.1 & 16,1,14,15,13,7 \\
J0621+1002 & 0.02885 & 36.56 & 17 & 58593 & 1.4 & 42.3 & 5.8 & 11,7,17 \\
J0629+2415 & 0.47662 & 84.18 & 14.5 & 58777 & 11.2 & 249.8 & 18 & 14,1,15,13,7,2 \\
J0636-3044 & 0.00395 & 15.46 & 7 & 56902 & 0.4 & 13.7 & 1 & 18 \\
J0646+0905 & 0.90391 & 149.7 & 10.6 & 58619 & 445.1 & 15 & 5.6 & 19,11 \\
J0647+0913 & 1.23485 & 154.7 & 12.2 & 58595 & 125.4 & 12.7 & 1.7 & 19,2 \\
J0658+2936 & 0.8237 & 39 & 19 & 56458 & 38.6 & 8.7 & 1 & 20,1 \\
J0711+0931 & 1.21409 & 46.24 & 13.9 & 58595 & 56.9 & 25.4 & 2.1 & 11,1,2 \\
J0854+5449 & 1.23303 & 18.84 & 12.7 & 55343 & 9.6 & 9.8 & 0.3 & 20 \\
J0927+2345 & 0.76189 & 17.24 & 6.1 & 58740 & 17.9 & 10.1 & 0.3 & 8,1 \\
J0943+2253 & 0.53297 & 27.25 & 10.2 & 58739 & 12.5 & 33.8 & 1.2 & 8,1,7 \\
J0946+0951 & 1.09771 & 15.32 & 13 & 58530 & 34.3 & 48.2 & 2.4 & 11,21,7 \\
J0953+0755 & 0.25307 & 2.97 & 15 & 58620 & 11.9 & 659.7 & 44.6 & 11,8,13,7,10,22 \\
J0957-0619 & 1.7237 & 26.95 & 7.7 & 59084 & 26.9 & 10.8 & 0.3 & 4 \\
J1000+08 & 0.44037 & 21 & 10.6 & 58595 & 13.8 & 82.9 & 3.4 & 18 \\
J1017+3011 & 0.45279 & 27.15 & 3.8 & 56053 & 21.2 & 11.3 & 0.5 & 7,2 \\
J1022+1001 & 0.01645 & 10.26 & 5.6 & 58465 & 1.7 & 222.8 & 14.8 & 8,13,7,22 \\
J1049+5822 & 0.72762 & 12.36 & 2.3 & 55290 & 22.7 & 8.1 & 0.3 & 23 \\
J1136+1551 & 1.18792 & 4.84 & 10.3 & 58876 & 37.1 & 1143.1 & 53.3 & 11,8,13,7,10 \\
J1238+2152 & 1.11859 & 17.97 & 17.3 & 58771 & 17.5 & 42.6 & 2.1 & 8,1,7 \\
J1246+2253 & 0.47387 & 17.79 & 13.8 & 58770 & 11.1 & 25.5 & 1.2 & 8,1,7 \\
J1313+0931 & 0.84893 & 12.04 & 19.4 & 58620 & 13.3 & 60.5 & 4.2 & 11,1,7 \\
J1334+1005 & 0.91109 & 23.87 & 2.3 & 58536 & 21.4 & 11.4 & 0.4 & 7 \\
J1343+6634 & 1.3941 & 30.03 & 8.6 & 55202 & 65.3 & 11.4 & 0.6 & 7 \\
J1404+1159 & 2.65044 & 18.5 & 12.8 & 58466 & 62.1 & 100.2 & 5.6 & 8,7,11,2 \\
J1513-2550 & 0.00212 & 46.88 & 4.6 & 56754 & 1 & 29.5 & 9.7 & 3 \\
J1520-19 & \nodata & \nodata & \nodata & 58624 & 46.3 & 11.7 & \nodata & \nodata \\
J1529+4050 & 0.47642 & 6.48 & 11.2 & 55689 & 70.7 & 14.5 & 1.7 & 11,7,2 \\
J1537+1155 & 0.0379 & 11.62 & 13.6 & 58872 & 1.8 & 21.7 & 2.4 & 8,1,13,7 \\
J1541+4703 & 0.2777 & 19.4 & 32.4 & 55436 & 136.9 & 9.2 & 17.6 & 20 \\
J1543+0929 & 0.74845 & 34.98 & 25.6 & 59083 & 52.6 & 131.1 & 46.9 & 11,13,7,10 \\
J1555-2908 & 0.00179 & 75.91 & 9.9 & 56907 & 0.3 & 13.4 & 2.6 & 3 \\
J1602-1009 & 0.00312 & 34.35 & 4.7 & 59086 & 0.2 & 23.9 & 2.1 & 11 \\
J1609-1930 & 1.55792 & 37 & 13.3 & 58773 & 24.3 & 31.3 & 1.8 & 24 \\
J1610-1322 & 1.01839 & 49.13 & 23.8 & 58532 & 71.6 & 29.4 & 8.7 & 21 \\
J1627+1419 & 0.49086 & 32.17 & 10.7 & 58872 & 34.5 & 29.7 & 4.6 & 8,1,7 \\
J1628-3205 & 0.00321 & 42.15 & 27.1 & 56933 & 0.2 & 11 & 7.1 & 3 \\
J1631+1252 & 0.31019 & 32.59 & 10.1 & 58895 & 4.8 & 8.5 & 0.5 & 9 \\
J1631-1612 & 0.67768 & 33.77 & 18.3 & 58767 & 31.8 & 15.9 & 2.4 & 4 \\
J1633-2009 & 0.93556 & 48.19 & 3.6 & 58775 & 43.9 & 34.2 & 3.3 & 4 \\
J1635-1511 & 1.17939 & 54 & 14.4 & 58742 & 165.8 & 8 & 1.7 & 24 \\
J1643+1338 & 1.09905 & 35.82 & 12.9 & 58873 & 25.8 & 28.9 & 2.7 & 11,1,7,2 \\
J1643-1224 & 0.00462 & 62.4 & 15.3 & 58593 & 1 & 54.4 & 16 & 12,24 \\
J1650-1654 & 1.74955 & 43.03 & 11.3 & 58769 & 54.7 & 29.8 & 2.5 & 12,11 \\
J1651-1709 & 0.9734 & 33.6 & 8.3 & 58770 & 30.4 & 92.7 & 7.1 & 10,21,15,11,12 \\
J1658+3630 & 0.03303 & 3.04 & 13.5 & 56072 & 7.2 & 14.7 & 2.5 & 7,1,2 \\
J1659-1305 & 0.64096 & 60.37 & 11 & 58593 & 15 & 44.4 & 3.3 & 21,12 \\
J1703-1846 & 0.80434 & 49.63 & 11.3 & 58772 & 25.1 & 132.6 & 13.7 & 25,11,10,12,24 \\
J1705-1906 & 0.29899 & 22.91 & 19.3 & 58772 & 9.3 & 122 & 22.4 & 10,26,18,11,12,24 \\
J1707+3556 & 0.15976 & 19.24 & 15.7 & 56072 & 5 & 16.1 & 1.1 & 7,1 \\
J1709-1640 & 0.65306 & 24.89 & 9.9 & 58769 & 15.3 & 433.6 & 35.6 & 25,21,15,18,11,10,12,24 \\
J1711-1509 & 0.86881 & 59.82 & 13.9 & 58624 & 20.4 & 49 & 4.7 & 21,18,11,12,24 \\
J1719-1438 & 0.00579 & 36.78 & 16.5 & 58622 & 0.9 & 12.2 & 4.3 & 18 \\
J1720-0533 & 0.00327 & 36.83 & 7.9 & 58252 & 0.3 & 16.7 & 2.5 & 9 \\
J1720-1633 & 1.56561 & 45.06 & 22.3 & 58764 & 36.7 & 58.7 & 12.3 & 15,18,11,12,24 \\
J1732-1930 & 0.48377 & 72.43 & 12.7 & 58773 & 15.1 & 28.4 & 4.6 & 12,18 \\
J1735+6320 & 0.51072 & 41.85 & 18 & 55249 & 8 & 9 & 0.6 & 7 \\
J1736+05 & 0.99925 & 42 & 15.2 & 58805 & 23.4 & 19.4 & 2.5 & 8,7,24,2 \\
J1738+0333 & 0.00585 & 33.77 & 9.8 & 58872 & 1.3 & 9.1 & 2.7 & 24,1,2 \\
J1741+1351 & \nodata & \nodata & \nodata & 58875 & 0.6 & 9.4 & \nodata & 24,26,1,2 \\
J1741-2019 & 3.90451 & 74.9 & 15.7 & 58776 & 183 & 24.5 & 7.2 & 24,18 \\
J1743-1351 & 0.40534 & 116.3 & 18.1 & 58622 & 12.7 & 23.9 & 5.1 & 15,18,11,12,24 \\
J1744-1134 & 0.00407 & 3.14 & 19.7 & 58388 & 0.2 & 35.4 & 10.8 & 12,26,18,24 \\
J1744-1610 & 1.75721 & 66.67 & 19 & 58764 & 27.5 & 23.3 & 4.2 & 24,18 \\
J1744-3922 & 0.17244 & 148.1 & 16.6 & 57538 & 18.7 & 15.3 & 7.1 & 27 \\
J1745-0059 & 0.67952 & 68.63 & 5.6 & 56774 & 21.2 & 51.5 & 4.9 & 28 \\
J1746-0156 & 1.82905 & 92.94 & 5.9 & 57542 & 57.1 & 8.5 & 0.8 & 28 \\
J1746-16 & \nodata & \nodata & \nodata & 58768 & 131.3 & 16.4 & \nodata & \nodata \\
J1748-1300 & 0.39413 & 99.36 & 15.5 & 58532 & 12.3 & 44.7 & 8.4 & 10,21,11,12,24 \\
J1749+5952 & 0.43604 & 45.07 & 9.9 & 55255 & 20.4 & 16 & 1.2 & 7 \\
J1753-1914 & 0.06295 & 105.3 & 18.1 & 58773 & 4.4 & 17 & 9 & 27,18 \\
J1755-1650 & 0.73374 & 159.9 & 16.6 & 58767 & 34.4 & 9.5 & 3 & 27,18 \\
J1759-1956 & 2.84339 & 236.4 & 13.3 & 58777 & 88.8 & 9.8 & 3.6 & 27,18 \\
J1801-1417 & 0.00363 & 57.25 & 8.2 & 58619 & 0.8 & 13.6 & 6.8 & 27,3 \\
J1805-1504 & 1.18127 & 225 & 12 & 58624 & 387.3 & 28 & 25.8 & 27,18 \\
J1806-1154 & 0.52262 & 122.23 & 14.8 & 58389 & 36.7 & 31.3 & 11.5 & 15,18,11,12,27 \\
J1807+04 & 0.7989 & 53 & 6.5 & 58867 & 37.4 & 10.1 & 1.2 & 1,2 \\
J1810+1744 & 0.00166 & 39.66 & 0.6 & 55325 & 0.5 & 226.1 & 57.8 & 3,4 \\
J1811+0702 & 0.46171 & 57.8 & 26.7 & 58805 & 21.6 & 8.4 & 3.7 & 8,7,24 \\
J1818+03\_P & 0.79919 & 98.7 & 5.3 & 58895 & 18.7 & 13.6 & 1.4 & 11 \\
J1819-1318 & 1.5157 & 35.1 & 10.7 & 58466 & 47.4 & 8.3 & 4.3 & 27 \\
J1822+0705 & 1.36282 & 62.2 & 30 & 58805 & 63.9 & 7.3 & 5.6 & 8,7,2 \\
J1822+2617 & 0.59142 & 64.7 & 12.2 & 56457 & 13.9 & 8.2 & 0.5 & 9 \\
J1823+0550 & 0.75291 & 66.78 & 13.8 & 58831 & 11.8 & 296.7 & 29.4 & 10,8,15,13,7,11,24,2 \\
J1824-1945 & 0.18934 & 224.26 & 4.5 & 58777 & 41.4 & 70.8 & 45.3 & 10,18,11,29,12,27 \\
J1826-1131 & 2.09314 & 319.4 & 21.1 & 58386 & 311.1 & 8.1 & 20.1 & 30,18,11,27 \\
J1829-1751 & 0.30714 & 216.83 & 10.8 & 58773 & 67.2 & 59.8 & 35.6 & 16,15,18,11,10,29,12,27 \\
J1834-1710 & 0.35831 & 123.8 & 3.3 & 58770 & 16.8 & 41 & 6.8 & 27,18 \\
J1834-1855 & 1.46566 & 185.2 & 16.2 & 58777 & 102.9 & 8.3 & 3.1 & 27,18 \\
J1835-1106 & 0.16592 & 132.68 & 2.2 & 58382 & 11.7 & 35.8 & 14.9 & 12,18,11,27 \\
J1837-1837 & 0.61836 & 100.96 & 1.9 & 58774 & 14.5 & 20.6 & 2.4 & 24,18,11,27 \\
J1839-1238 & 1.91143 & 173.85 & 17.9 & 58383 & 417.7 & 14.9 & 14.9 & 27,18,11 \\
J1840+2843 & 0.6861 & 63.7 & 25.8 & 56180 & 676.5 & 19.8 & 247.7 & 11 \\
J1840-0815 & 1.09644 & 233.75 & 9.1 & 59085 & 239.6 & 16 & 15.4 & 27,18,11 \\
J1840-1207 & 0.75447 & 290.9 & 10.5 & 58383 & 35.4 & 11.9 & 3.1 & 27,18 \\
J1843-1113 & 0.00185 & 59.96 & 15 & 58382 & 0.4 & 15.6 & 11.9 & 27,26,18 \\
J1845+0623 & 1.42165 & 115.72 & 3.9 & 58805 & 22.2 & 15.5 & 1.5 & 27,11,2 \\
J1848-1150 & 1.31222 & 163.4 & 18.8 & 58383 & 41 & 7.5 & 2.4 & 27,18 \\
J1848-1414 & 0.29777 & 134.47 & 14.9 & 58387 & 14 & 11.9 & 2.8 & 12,18,24 \\
J1848-1952 & 4.30819 & 18.23 & 14.9 & 58777 & 67.3 & 139 & 15.6 & 10,18,24 \\
J1851+0418 & 0.2847 & 115.34 & 14.5 & 59085 & 62.3 & 14 & 11.8 & 11,17,27,2 \\
J1854-1421 & 1.14659 & 130.4 & 2.9 & 58470 & 26.9 & 56.6 & 5.3 & 10,18,12,24 \\
J1854-1557 & 3.45312 & 150 & 24.1 & 58742 & 107.9 & 11.6 & 3.9 & 18 \\
J1858-11 & \nodata & \nodata & \nodata & 58383 & 42 & 12.8 & \nodata & \nodata \\
J1900+5106 & 0.33776 & 71.6 & 12.1 & 55417 & 7.9 & 10.3 & 0.6 & 31 \\
J1901-04 & \nodata & \nodata & \nodata & 57544 & 271.1 & 16.1 & \nodata & \nodata \\
J1902+0556 & 0.74658 & 177.49 & 13.9 & 58805 & 52.5 & 24.2 & 16.6 & 32,8,30,17,27 \\
J1902+0615 & 0.67351 & 502.93 & 17.3 & 58805 & 220.8 & 7.9 & 16.1 & 30,11,17,27,2 \\
J1904-1224 & 0.75081 & 118.28 & 13.3 & 58383 & 23.5 & 33.8 & 4.8 & 12,18,11,24 \\
J1904-1629 & 1.54141 & 145 & 12.4 & 58768 & 36.1 & 9.7 & 1 & 18 \\
J1904-18 & \nodata & \nodata & \nodata & 58773 & 79.1 & 13.1 & \nodata & \nodata \\
J1907-1532 & 0.63224 & 72.6 & 7.1 & 58741 & 19.8 & 14.3 & 1.3 & 18 \\
J1908+0500 & 0.29102 & 201.24 & 15.5 & 59085 & 13.6 & 10.2 & 4.7 & 33,11,17,27,2 \\
J1910+0358 & 2.33027 & 83.87 & 27.1 & 59085 & 509.8 & 19.1 & 43.3 & 11,7,10,27 \\
J1913+0446 & 1.61613 & 109.1 & 13.1 & 59085 & 25.3 & 15 & 3.3 & 27,2 \\
J1919+2621 & 0.65151 & 96.5 & 7.1 & 56159 & 15.3 & 8.1 & 0.5 & 9,2 \\
J1924+2040 & 0.23779 & 226.6 & 13.6 & 56480 & 11.1 & 10 & 1.8 & 32,2 \\
J1926+0431 & 1.07408 & 102.24 & 9.5 & 59085 & 25.2 & 63.5 & 7.7 & 10,13,7,24,2 \\
J1930-1852 & 0.18552 & 42.85 & 0 & 57473 & 5.8 & 67.8 & 4.1 & 4 \\
J1938+0650 & 1.12156 & 70.8 & 26.2 & 58777 & 26.3 & 10.7 & 3.9 & 8,7,24,2 \\
J1938+2659 & 0.88333 & 191.2 & 12.8 & 56158 & 41.4 & 8 & 1.2 & 17,2 \\
J1938-0940 & 0.17387 & 44.47 & 14.3 & 58174 & 4.1 & 13.4 & 1.3 & 9 \\
J1940-0902 & 0.97847 & 42.3 & 9.3 & 59081 & 30.6 & 14.3 & 1.2 & 34 \\
J1942+3941 & 1.35329 & 104.5 & 14.4 & 55851 & 95.2 & 11.2 & 1.6 & 9,2 \\
J1942-2019 & 0.27513 & 59.9 & 13.1 & 58775 & 6.4 & 20.8 & 1.4 & 34 \\
J1943+0609 & 0.44623 & 70.66 & 15.6 & 58777 & 20.9 & 22.2 & 3.9 & 24,8,7,11,2 \\
J1944-10 & 0.40914 & 31.01 & 11.4 & 58233 & 9.6 & 8.7 & 0.6 & 4 \\
J1944-1750 & 0.84116 & 56.32 & 11.2 & 58769 & 26.3 & 114 & 7.8 & 10,12,24 \\
J1952+3021 & 1.66567 & 189.1 & 13.3 & 56082 & 39 & 10.8 & 1.3 & 17,1,7,2 \\
J1955+2930 & 1.07388 & 212.4 & 3.6 & 56086 & 33.6 & 12.8 & 1.3 & 17,1,2 \\
J2006+4058 & 0.49969 & 259.5 & 16.3 & 55689 & 74.2 & 9.1 & 4.4 & 9,2,35 \\
J2024+48 & 1.262 & 99 & 42.9 & 55440 & 39.4 & 10 & 46.8 & 4 \\
J2027+0255 & 0.0106 & 51.13 & 11.9 & 58769 & 1.6 & 9.6 & 1.5 & 9 \\
J2038+3447 & 0.16016 & 57.93 & 20.7 & 56046 & 3.8 & 8.5 & 1.4 & 11 \\
J2043+31 & 0.937 & 145 & 4.3 & 56079 & 29.2 & 25 & 1.8 & 6 \\
J2045+0912 & 0.39556 & 31.78 & 11.4 & 58774 & 18.5 & 59.5 & 4.5 & 8,7 \\
J2045+3633 & 0.03168 & 129.53 & 5.8 & 55991 & 15.6 & 13.6 & 5.9 & 36,1,2 \\
J2047+1053 & 0.00429 & 34.63 & 3.1 & 58773 & 0.6 & 11.6 & 1.3 & 3,1 \\
J2047+5029 & 0.44594 & 107.68 & 8.7 & 55441 & 15.7 & 8.4 & 1.7 & 11,2 \\
J2050+1259 & 1.22102 & 52.4 & 11.3 & 58621 & 181.3 & 8.9 & 1.4 & 11 \\
J2051+1248 & 0.55317 & 43.45 & 12.8 & 58621 & 56.2 & 33.3 & 4.4 & 11,7,2 \\
J2057+4701 & 0.55957 & 219 & 10.7 & 55441 & 26.2 & 10.1 & 1.5 & 20,35 \\
J2100+4712\_P & 1.45872 & 233.7 & 24.1 & 55558 & 717.2 & 8.5 & 14.4 & 2 \\
J2112+4058 & 4.06075 & 129 & 9.4 & 55848 & 95.3 & 13.4 & 1 & 9,2 \\
J2115+6702 & 0.55212 & 54.5 & 8.6 & 56655 & 25.9 & 22.6 & 1.7 & 11 \\
J2116+1414 & 0.44015 & 56.2 & 22.9 & 58619 & 13.8 & 34.9 & 4.8 & 10,7,11 \\
J2116+3701 & 0.14588 & 44.03 & 15.1 & 55991 & 2.3 & 10 & 0.6 & 20,1 \\
J2122+2426 & 0.54142 & 8.5 & 12.6 & 56254 & 16.9 & 9.2 & 0.6 & 7,2 \\
J2129+4119 & 1.68742 & 32 & 18.6 & 55847 & 118.7 & 10.4 & 1.9 & 9,2 \\
J2208+4610 & 0.64251 & 63.41 & 11.8 & 55557 & 65.3 & 10.3 & 1.5 & 20,2 \\
J2215+5135 & 0.00261 & 69.19 & 3.4 & 55346 & 0.4 & 71.7 & 10.6 & 3,4,2 \\
J2226-03 & 0.7697 & 17.5 & 12.1 & 57211 & 36.1 & 14.2 & 0.8 & 6 \\
J2229+40 & 0.2729 & 74 & 15.3 & 55847 & 12.8 & 10.8 & 1.2 & 6 \\
J2251-3711 & 12.12256 & 12.12 & 25.9 & 57062 & 1797.7 & 8.9 & 3.3 & 34 \\
J2302+48 & 0.7421 & 72 & 13.2 & 55557 & 23.2 & 13 & 1 & 6 \\
J2306+3124 & 0.3416 & 46.15 & 17.1 & 56078 & 10.7 & 7.2 & 0.6 & 7,1,2 \\
J2336-0151 & 1.02984 & 19.62 & 18.3 & 59082 & 32.2 & 20.8 & 1.6 & 7 \\
J2340+08 & 0.3033 & 24 & 11.5 & 58705 & 14.2 & 25.2 & 1.6 & 1,7,2 \\
\enddata
\tablecomments{The Survey column lists numerical references corresponding to the comma-separated survey labels in the ATNF \texttt{SURVEY} field. Pulsars associated with more than one listed survey have multiple reference numbers. The angular separation $\theta$ is between the catalog pulsar position and the GBNCC pointing center. The quoted $S_{350}$ values are the flux-density estimates derived from the corresponding GBNCC observations. The ATNF \texttt{misc} label is a catch-all category and does not correspond to a single survey publication. A missing or dash-valued SURVEY entry is written as \texttt{\string\nodata}.}
\tablerefs{
1: AO327---\citet{2013ApJ...775...51D};
2: FAST GPPS---\citet{2021RAA....21..107H};
3: Fermi Pulsar Search Consortium---\citet{2012arXiv1205.3089R};
4: GBT 350-MHz Drift-scan Survey---\citet{2013ApJ...763...80B};
5: SMART---\citet{2023PASA...40...21B};
6: Pushchino Multibeam Pulsar Search---\citet{2022PoS...425...43T};
7: LOTAAS---\citet{sanidas_19};
8: Arecibo Survey 4---\citet{1995ApJ...454..826F};
9: FAST CRAFTS---\citet{2018IMMag..19..112L};
10: Second Molonglo Pulsar Survey---\citet{1978MNRAS.185..409M};
11: ATNF misc---no single survey reference;
12: Parkes Southern Pulsar Survey---\citet{1996MNRAS.279.1235M};
13: Green Bank Survey 4---\citet{1997ApJ...474..426S};
14: Green Bank Survey 1---\citet{1978ApJ...225L..31D};
15: Green Bank Survey 3---\citet{1985Natur.317..787S};
16: Jodrell Survey A---\citet{1977MNRAS.179..635D};
17: PALFA---\citet{2006ApJ...637..446C};
18: HTRU-South---\citet{2010MNRAS.409..619K};
19: Parkes Perseus Arm Pulsar Survey---\citet{2013MNRAS.429..579B};
20: CHIME pulsar survey---\citet{2021ApJ...922...43G};
21: Green Bank Survey 2---\citet{1985ApJ...294L..25D};
22: Parkes High-Latitude Pulsar Survey---\citet{2006MNRAS.368..283B};
23: TULIPP---\citet{sobey_22};
24: Swinburne Intermediate-Latitude Pulsar Survey---\citet{2001MNRAS.326..358E};
25: First Molonglo Pulsar Survey---\citet{1968Natur.220..753L};
26: Fermi LAT blind pulsar searches---\citet{2009Sci...325..840A};
27: Parkes Multibeam Pulsar Survey---\citet{2001MNRAS.328...17M};
28: FAST high-Galactic-latitude pulsar survey---\citet{2025ApJ...982..117X};
29: Parkes 20-cm survey---\citet{1992MNRAS.255..401J};
30: Jodrell Survey B---\citet{1986Natur.320...43C};
31: CHAMPSS---\citet{2025ApJ...990...50A};
32: Arecibo Survey 1---\citet{1974ApJ...191L..59H};
33: Arecibo Survey 3---\citet{1995ApJ...449..156N};
34: SUPERB---\citet{2018MNRAS.473..116K};
35: Green Bank 820-MHz Pulsar Survey---\citet{2024ApJ...969..118M};
36: HTRU-North---\citet{2013MNRAS.435.2234B}.}
\end{deluxetable*}

\section{Appendix information}


\bibliography{sample701}{}
\bibliographystyle{aasjournalv7}



\end{document}